\documentclass[aps,twocolumn,prl,notitlepage,floats,superscriptaddress,amsmath,amssymb]{revtex4-2}

\usepackage[version=3]{mhchem} 
\usepackage{bm}
\usepackage[utf8]{inputenc}
\usepackage[T1]{fontenc}
\usepackage{graphicx}
\usepackage{upgreek}
\usepackage{color}
\usepackage{ulem}
\usepackage{amsmath}
\usepackage{hyperref}
\hypersetup{colorlinks,citecolor=blue, filecolor=blue ,linkcolor=blue , urlcolor=blue, pdftex}
\usepackage{sidecap}
\usepackage{amssymb}
\usepackage[caption=false]{subfig}
\usepackage{orcidlink}
\usepackage[dvipsnames]{xcolor}
\usepackage{multirow}
\usepackage[dvipsnames]{xcolor}
\usepackage{array}
\usepackage{xr}

\setcitestyle{super}

\def \FUW{Faculty of Physics, University of Warsaw, Pasteura 5, 02-093 Warsaw, Poland}
\def \SingaporeEng{Department of Materials Science and Engineering, National University of Singapore, 9 Engineering Dr 1, 117575, Singapore} 
\def \SingaporeInt{Institute for Functional Intelligent Materials, National University of Singapore, 9 Engineering Dr 1, 117544, Singapore} 
\def \Zdenek {Department of Inorganic Chemistry, University of Chemistry and Technology, Technická 5, 160 00 Praha 6-Dejvice, Prague, Czech Republic}
\def \PWr {Faculty of Fundamental Problems of Technology, Wroc\l{}aw Univwersity of Science and Technology, Wyb. Wyspia{\'n}skiego 27, 50-370 Wroc\l{}aw, Poland}

\begin{document}
\title{Strong Spin-Lattice Interaction in Layered Antiferromagnetic CrCl$_\textrm{3}$}

\author{\L{}ucja Kipczak\orcidlink{0000-0003-1266-0201}}
\email{lucja.kipczak@fuw.edu.pl}
\affiliation{\FUW} 
\author{Tomasz Wo\'zniak}
\affiliation{\FUW} 
\affiliation{\PWr}
\author{Chinmay K. Mohanty}
\affiliation{\FUW}
\author{Igor Antoniazzi}
\affiliation{\FUW}
\author{Jakub~Iwa\'nski\orcidlink{0000-0003-2395-4010}}
\affiliation{\FUW}
\author{Przemys\l{}aw~Oliwa\orcidlink{0000-0003-1255-5997}}
\affiliation{\FUW}
\author{Jan Paw\l{}owski}
\affiliation{\FUW}
\author{Meganathan Kalaiarasan}
\affiliation{\Zdenek}
\author{Zden\v{e}k~Sofer}
\affiliation{\Zdenek}
\author{Andrzej~Wysmo\l{}ek\orcidlink{0000-0002-8302-2189}}
\affiliation{\FUW}
\author{Adam Babi\'nski\orcidlink{0000-0002-5591-4825}}
\affiliation{\FUW} 
\author{Maciej Koperski\orcidlink{0000-0002-8301-914X}}
\affiliation{\SingaporeInt}
\affiliation{\SingaporeEng}
\author{Maciej R. Molas\orcidlink{0000-0002-5516-9415}}
\email{maciej.molas@fuw.edu.pl}
\affiliation{\FUW}

\begin{abstract}
Understanding the coupling between lattice vibrations and magnetic order is crucial for controlling properties of two-dimensional magnetic materials. 
Here, we investigate the vibrational properties of bulk and thick-flake CrCl$_\textrm{3}$ using polarization-resolved Raman spectroscopy, complemented by photoluminescence, photoluminescence excitation, and optical absorption measurements. 
Symmetry analysis, supported by first-principles phonon calculations, enables the unambiguous assignment of all eight Raman-active modes, four $\textrm{A}_\textrm{g}$ and four $\textrm{E}_\textrm{g}$, previously predicted only theoretically. 
Excitation-energy-dependent measurements reveal that the strong enhancement of selected phonon modes originates primarily from interference effects rather than resonant Raman scattering.
Temperature-dependent Raman spectroscopy further reveals pronounced signatures of spin-phonon coupling across the transition from a fully antiferromagnetic phase, through an intermediate regime with local, domain-like ferromagnetic order, to the paramagnetic phase, accompanied by a clear rhombohedral-to-monoclinic structural transition.
Together, these results demonstrate how lattice, electronic, and magnetic degrees of freedom collectively govern the Raman response of CrCl$_\textrm{3}$.
\end{abstract}

\maketitle

The discovery of a long-range magnetism in two-dimensional (2D) van der Waals (vdW) materials has opened new opportunities to explore magnetic phenomena in reduced dimensionality and to integrate magnetic order into layered heterostructures. 
Furthermore, interplay between sample thickness, external magnetic fields, and optical excitations governs the stabilization of the magnetic ordering, including the formation of topological spin textures.\cite{Grebenchuk_2024,Grebenchuk_2024_JPM,Fullerton_2025}
These advances substantially broaden the scope of spintronic,\cite{Mi_2023, Jia2025} valleytronic,\cite{Luo_2024} and quantum magneto-optical applications.\cite{Jiang2018, Huang2018, Gilbertini2019, Soriano2020, Wang2022}

Chromium trihalides, with the chemical formula CrX$_\textrm{3}$ (X = Cl, Br, I), constitute a prototypical family of vdW magnetic materials.
In bulk CrX$_\textrm{3}$, strong intralayer ferromagnetic (FM) coupling is present within each \mbox{X-Cr-X} layer; however, both the magnetic anisotropy and the interlayer exchange interactions vary markedly across the halide series.
In particular, CrCl$_\textrm{3}$ exhibits antiferromagnetic (AFM) interlayer coupling and an in-plane easy axis,\cite{McGuire2017Crystals, McGuire_2017, Mak2019, Serri2020} in sharp contrast to the heavier chromium trihalides.
On the other hand, CrBr$_\textrm{3}$ displays FM interlayer coupling and strong out-of-plane magnetic anisotropy, resulting in spins aligned perpendicular to the layers,\cite{Dillon1963, Dillion1966} whereas CrI$_\textrm{3}$, despite its pronounced out-of-plane anisotropy arising from strong spin-orbit coupling, exhibits AFM interlayer coupling in the bilayer and few-layer limits.\cite{McGuire_2017, McGuire2015, Li2020}
Moreover, the interlayer magnetic coupling is highly sensitive to the stacking order. 
Stacking faults commonly occur in these materials due to the small energy differences between competing crystallographic structures, and such structural variations can strongly impact the magnetic order.\cite{Grebenchuk_2025}

Beyond their intrinsic magnetic and structural properties, CrX$_\textrm{3}$ compounds are well suited for vdW heterostructures, in which magnetic layers are combined with nonmagnetic semiconductors or other 2D materials.
In such hybrid systems, interfacial proximity effects enable the transfer of magnetic order, exchange fields, or spin textures to adjacent layers.\cite{Heissenbuttel2021, Behera2019, Ciorciaro2020, Kipczak2025, Jana_2025}
These engineered heterostructures provide a versatile platform for the realization of novel spin-based and magneto-optical devices.

In this work, we investigate spin-phonon interactions in bulk CrCl$_\textrm{3}$ and thick exfoliated flakes using Raman scattering (RS) and complementary optical spectroscopies.
Polarization-resolved RS measurements, combined with first-principles phonon calculations, enable the symmetry assignment of all eight Raman-active phonon modes, four $\textrm{A}_\textrm{g}$ and four $\textrm{E}_\textrm{g}$ modes, completing their experimental verification beyond theoretical predictions.
Excitation-energy-dependent RS and optical spectroscopy show that interference effects dominate the enhancement of the Raman response.
Temperature-dependent Raman measurements uncover strong spin-phonon coupling and reveal pronounced anomalies at the magnetic ordering temperature, together with a clear rhombohedral-to-monoclinic structural transition.


\begin{figure*}[!th]
		\subfloat{}%
		\centering
		\includegraphics[width=1\linewidth]{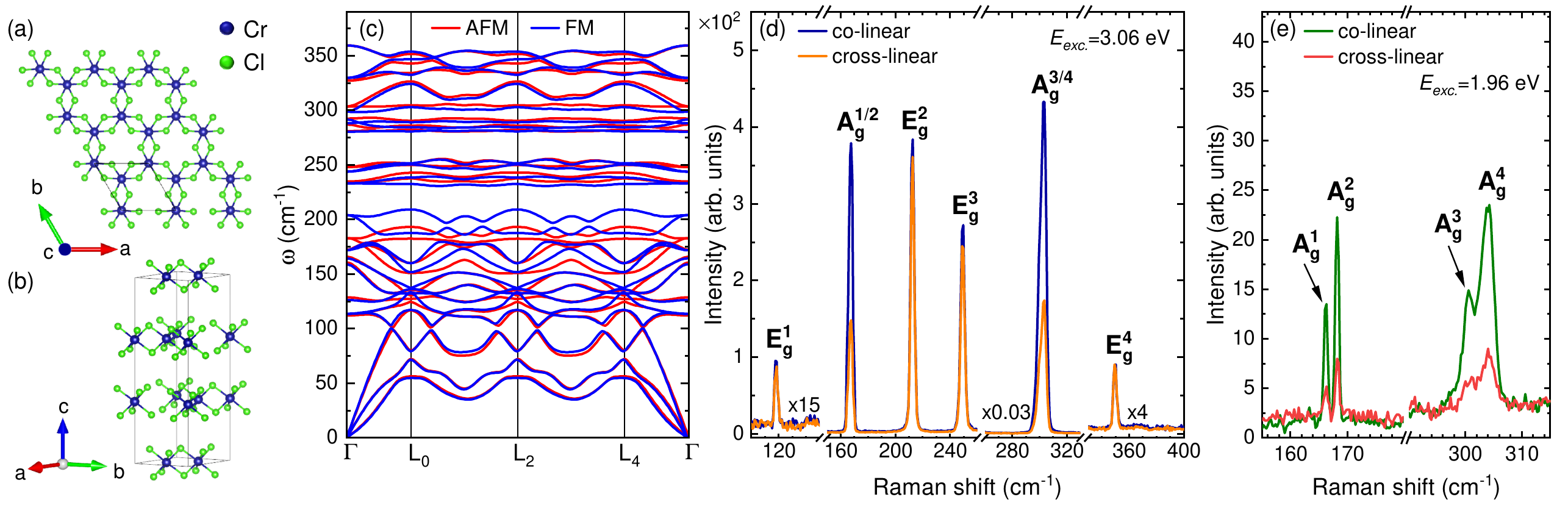}
      	\caption{Schematic representation of the rhombohedral crystal structure of CrCl$_\textrm{3}$.
        (a) Top view of a monolayer, with the unit cell indicated by a black rectangle and shown in a simplified form for clarity.
        (b) Perspective view of the bulk structure, illustrating the stacking of CrCl$_\textrm{3}$ layers along the \textit{c}-axis and the arrangement of multiple layers within the unit cell.
        (c) Phonon dispersion of bulk CrCl$_\textrm{3}$ calculated for the rhombohedral primitive cell with ferromagnetic (bluFe) and antiferromagnetic (red) spin configurations. 
        (d) and (e) Low-temperature ($T$=5~K) Raman scattering spectra of bulk CrCl$_\textrm{3}$ measured in co-linear and cross-linear polarization configurations using 3.06 eV and 1.96~eV laser excitation, respectively, with a laser power of 75~$\mu$W. 
        Panel (e) is limited to two $\textrm{A}_\textrm{g}$ modes exhibiting doublet structures.
        }  
		\label{fig:structure}
\end{figure*}

CrCl$_\textrm{3}$ crystallizes in a monoclinic structure with $C2/m$ symmetry at ambient temperature and undergoes a transition to a rhombohedral phase upon cooling.\cite{McGuire_2017}
At low temperature ($T\sim$10~K), CrCl$_\textrm{3}$ adopts $R\bar{3}$ symmetry, corresponding to a primitive rhombohedral unit cell belonging to space group No.~148.\cite{Cable1961, Kuhlow1982, KANESAKA1986, McGuire2017Crystals}
Figures~\ref{fig:structure}(a)-(c) illustrate the crystal structure of the rhombohedral CrCl$_\textrm{3}$, including a top view of a monolayer (a) and the bulk stacking sequence (b).
Within each layer, Cr atoms form a hexagonal lattice, with each Cr atom octahedrally coordinated by six chlorine (Cl) atoms through strong covalent bonding.
Adjacent layers are coupled by weak vdW interactions,\cite{McGuire2017Crystals, Gilbertini2019} which enable mechanical exfoliation.\cite{Liu2016PCCP, McGuire_2017, Webster2018}
Phonon dispersions, shown in Fig.~\ref{fig:structure}(c), were calculated for both the AFM and FM phases using density functional theory (DFT) in light of recent reports of coexisting magnetic orders in bulk CrCl$_\textrm{3}$.\cite{Schneeloch_2024}
The computational details are provided in the Methods section of the Supporting Information (SI).
The calculated lattice constant $a$=6.7069~$\text{\AA}$ and rhombohedral angle $\alpha$=52.9688$^\circ$ agree well with the experimentally reported values of $a$=6.7194~$\text{\AA}$ and $\alpha$=52.4825$^\circ$ measured at $T$=225~K.\cite{Morosin1964}
At the $\Gamma$ point, eight Raman-active phonon modes are obtained for both AFM and FM orders, which can be classified according to the irreducible representation $\Gamma{_\textrm{AFM/FM}} = 4\textrm{A}_\textrm{g} + 4\textrm{E}_\textrm{g}$.

We investigated the phonon modes of bulk CrCl$_\textrm{3}$, $i.e.$, a macroscopic single crystal, as described in the Methods section of the SI.
Polarization-resolved low-temperature ($T$=5~K) RS spectra of bulk CrCl$_\textrm{3}$, measured using 3.06~eV excitation, are presented in Fig.~\ref{fig:structure}(d).
Six Raman-active modes are experimentally resolved, consisting of two modes with $\textrm{A}_{\textrm{g}}$ symmetry and four modes with $\textrm{E}_{\textrm{g}}$ symmetry.
The numerical superscripts assigned to the modes serve solely as labels and do not denote their symmetry character.

The mode assignments were established through polarization-resolved RS measurements performed in co-linear and cross-linear configurations.
As shown in the figure, the intensities of the $\textrm{E}_{\textrm{g}}$ modes remain essentially unchanged between the two polarization geometries, whereas the $\textrm{A}_{\textrm{g}}$ modes are strongly suppressed, by approximately a factor of three, in the cross-linear configuration.
Because the $\textrm{A}_{\textrm{g}}$ symmetry peaks exhibit asymmetric line shapes under 3.06~eV excitation, additional polarization-resolved RS measurements were carried out using 1.96~eV excitation to achieve improved spectral resolution (Fig.~\ref{fig:structure}(e)).
These measurements clearly demonstrate that both the $\textrm{A}^{1/2}_{\textrm{g}}$ and $\textrm{A}^{3/4}_{\textrm{g}}$ features consist of two well-resolved peaks, each exhibiting $\textrm{A}_{\textrm{g}}$ symmetry.
To rule out sample-quality effects related to crystal growth, complementary measurements were performed on a commercially available CrCl$_\textrm{3}$ crystal, which displays an identical double-peak structure for these modes (see Section S1 of the SI).
Experimental artifacts can also be excluded, as the double-peak structure is observed exclusively for the $\textrm{A}^{1/2}_{\textrm{g}}$ and $\textrm{A}^{3/4}_{\textrm{g}}$ modes, whereas all other Raman-active modes consistently exhibit single, well-defined peaks.

These results establish that the low-temperature RS spectra of bulk CrCl$_\textrm{3}$ comprise eight phonon modes, $i.e.$, four $\textrm{A}_{\textrm{g}}$ and four $\textrm{E}_{\textrm{g}}$ modes, in full agreement with theoretical predictions for both AFM and FM orders.
Table~\ref{table:1} compares the calculated phonon frequencies with the experimentally measured Raman modes and shows excellent overall agreement.
While the polarization behavior of the $\textrm{E}_{\textrm{g}}$ modes follows symmetry-based selection rules, the $\textrm{A}_{\textrm{g}}$ modes are expected to vanish entirely in the cross-linear configuration.\cite{Klein2019, Huang2020, Mai2021}
The finite $\textrm{A}_{\textrm{g}}$ intensities observed under cross-linear polarization are attributed to resonant excitation effects, as previously reported for thin layers of semiconducting transition metal dichalcogenides.\cite{Lee2018-tmd-pol, Tan2021}
Angle-resolved RS measurements, presented in Section S2 of the SI, further confirm the isotropic in-plane symmetry of bulk CrCl$_\textrm{3}$.

\begin{table}[b]
\centering
\caption{Comparison of vibrational mode frequencies observed experimentally and calculated using DFT. }
\label{table:1}
\renewcommand{\arraystretch}{1.5}
\begin{tabular}{ c || c | c | c}
\multirow{ 2}{*}{\textbf{Mode}} & \multirow{ 2}{*}{\textbf{Experiment}} & \multicolumn{2}{c}{\textbf{DFT calculations}} \\ 
  &  & \textbf{FM phase} & \textbf{AFM phase}\\
 \hline \hline
    $\textrm{E}^{1}_\textrm{g}$   &  118.7 cm$^{-1}$  &  113.5 cm$^{-1}$  &  111.9 cm$^{-1}$\\ 
    $\textrm{A}^{1}_\textrm{g}$   &  166.1 cm$^{-1}$  &  158.7 cm$^{-1}$  &  160.5 cm$^{-1}$\\ 
    $\textrm{A}^{2}_\textrm{g}$   &  168.2 cm$^{-1}$  &  164.2 cm$^{-1}$  &  165.7 cm$^{-1}$\\ 
    $\textrm{E}^{2}_\textrm{g}$   &  212.8 cm$^{-1}$  & 203.8 cm$^{-1}$ & 182.8 cm$^{-1}$ \\ 
    $\textrm{E}^{3}_\textrm{g}$   & 249.5 cm$^{-1}$  & 233.0 cm$^{-1}$ & 234.6 cm$^{-1}$\\
    $\textrm{A}^{3}_\textrm{g}$   &  300.5 cm$^{-1}$  &  280.3 cm$^{-1}$  &  280.8 cm$^{-1}$\\ 
    $\textrm{A}^{4}_\textrm{g}$   &  303.9 cm$^{-1}$  &  281.2 cm$^{-1}$  &  286.9 cm$^{-1}$\\ 
    $\textrm{E}^{4}_\textrm{g}$   & 350.3 cm$^{-1}$ & 329.7 cm$^{-1}$ & 327.0 cm$^{-1}$ \\
\end{tabular}
\end{table}


\begin{figure*}[!th]
		\subfloat{}%
		\centering
		\includegraphics[width=1 \linewidth]{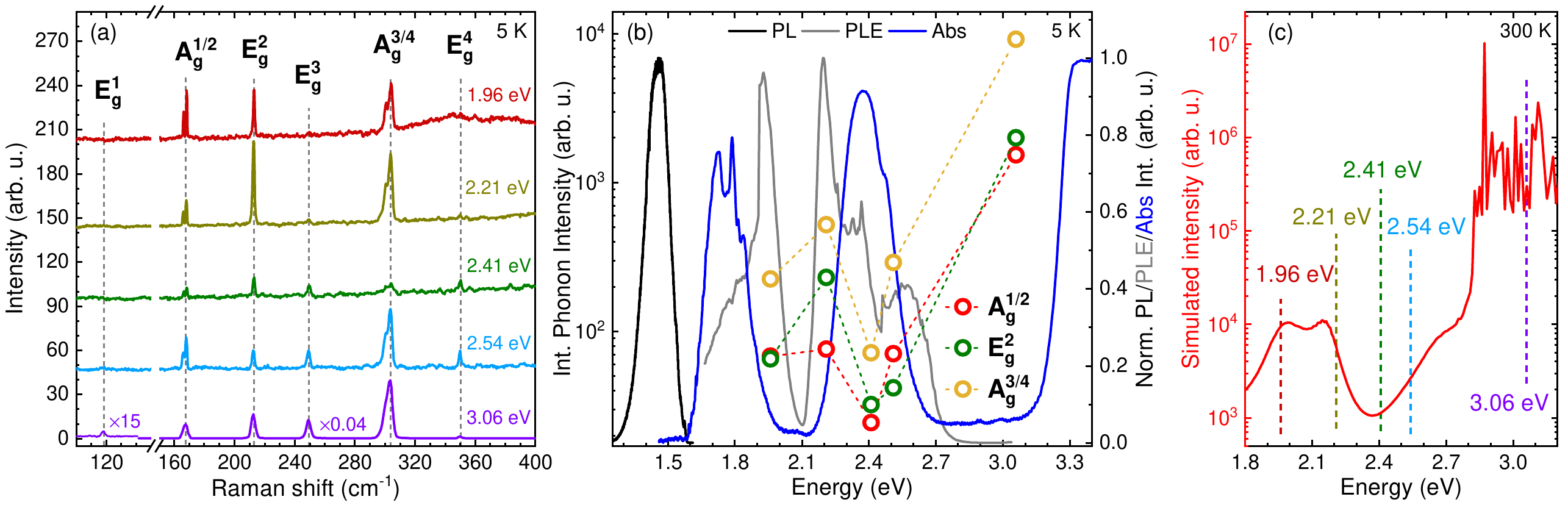}
   	\caption{Resonant Raman scattering investigation in the CrCl$_\textrm{3}$ crystal.
    (a) Raman scattering spectra of the CrCl$_\textrm{3}$ crystal measured at 5~K using different excitation energies: 1.96 eV, 2.21 eV, 2.41 eV, 2.54 eV, 3.06 eV, with an excitation power of 50 $\mu$W. 
    The spectra have been vertically shifted for better visual clarity. 
    (b) Left axis: graph shows points representing the intensity of individual Raman modes present in every spectra collected with different excitation lasers (in logarithmic scale).
    Right axis: normalized PL (black), PLE (grey) and Abs (blue) spectra measured on the CrCl$_{3}$ crystal. 
    The PL spectrum was measured with 2.21 eV laser with power around 0.5 $\mu$W.
    The PLE spectrum was detected via supercontinuum laser with power around 2 $\mu$W.
    (c) Simulated enhancement of the A$^{3/4}_\mathrm{g}$ intensity using the transfer-matrix method. 
    The colored vertical dashed lines indicate the excitation energies used in the experiment.}
    \label{fig:raman-interference}
\end{figure*}

To evaluate the influence of excitation energy on the RS spectra, we measured low-temperature RS spectra of bulk CrCl$_\textrm{3}$ using multiple excitation energies, $i.e.$, 3.06 eV, 2.54 eV, 2.41 eV, 2.21 eV, and 1.96~eV, as shown in Fig.~\ref{fig:raman-interference}(a).
The Raman modes intensities are comparable for most excitation energies, whereas a pronounced enhancement is observed under the 3.06~eV excitation.
For clarity, the spectrum acquired at 3.06~eV is scaled by a factor of 0.04, and the spectral region between 100~cm$^{-1}$ and 130~cm$^{-1}$ is further magnified by a factor of 15 to highlight all Raman features.
Across all excitation energies, six Raman-active modes are observed, consistent with previous reports on CrCl$_\textrm{3}$ crystals.\cite{Bermudez1976, KANESAKA1986, Kazim_2020}
To clarify the role of the excitation energy, the behavior of individual Raman modes is analyzed below.

The $\textrm{E}^{1}_{\textrm{g}}$ mode is detected exclusively under 3.06~eV excitation and remains substantially weaker than the other modes.
The $\textrm{E}^{2}_{\textrm{g}}$ mode, together with the $\textrm{A}^{1/2}_{\textrm{g}}$ and $\textrm{A}^{3/4}_{\textrm{g}}$ doublets, exhibits a non-monotonic dependence on excitation energy (discussed in detail below) and reaches maximum relative intensity at 2.21~eV.
In contrast, the $\textrm{E}^{3}_{\textrm{g}}$ and $\textrm{E}^{4}_{\textrm{g}}$ modes display nearly identical excitation-energy dependences: their intensities are strongly suppressed as the excitation energy decreases from 3.06 to 2.41~eV, become barely discernible at 2.21~eV, and vanish entirely at 1.96~eV.

To probe the origin of the resonant Raman response in CrCl$_\textrm{3}$, we extracted the integrated intensities of the $\textrm{E}^{2}_{\textrm{g}}$, $\textrm{A}^{1/2}_{\textrm{g}}$, and $\textrm{A}^{3/4}_{\textrm{g}}$ modes by Lorentzian deconvolution of the spectra for all excitation energies, as shown in Fig.~\ref{fig:raman-interference}(b).
The intensities of the $\textrm{A}^{1/2}_{\textrm{g}}$ and $\textrm{A}^{3/4}_{\textrm{g}}$ modes are treated as a combined doublet intensity, since their individual contributions cannot be reliably deconvoluted under 3.06~eV excitation.
The extracted phonon intensities span nearly two orders of magnitude, underscoring the presence of strong enhancement effects.

The intensity of the RS signal in vdW materials is known to depend sensitively on the excitation energy.\cite{Carvalho2015, Zhang2015, Grzeszczyk2016, Miranda2017}
Within a simple picture, resonant enhancement of RS, governed by electron-phonon coupling, occurs when the excitation energy approaches an electronic or excitonic transition of the material.\cite{Carvalho2015}
To identify transitions relevant to the resonant conditions of RS in CrCl$_\textrm{3}$, we measured the photoluminescence excitation (PLE) spectrum by monitoring the broad photoluminescence (PL) band of bulk CrCl$_\textrm{3}$ while continuously tuning the excitation energy, as shown in Fig.~\ref{fig:raman-interference}(b).
The PL spectrum of bulk CrCl$_\textrm{3}$ is characteristic of a molecular crystal, with Frenkel-type excitons localized on individual molecules and radiative recombination governed by the Franck-Condon principle.\cite{Acharya2022,refractive-index, CrBr3_spin_pumping}
The PLE spectrum reveals two broad excitation bands centered at approximately 1.89~eV and 2.38~eV.
The lower-energy band consists of two resonances at about 1.79~eV and 1.94~eV, while the higher-energy band comprises three resonances at approximately 2.20~eV, 2.34~eV, and 2.56~eV.
We further measured the absorbance (Abs) spectrum of a CrCl$_\textrm{3}$ crystal, shown in Fig.~\ref{fig:raman-interference}(b).
The Abs spectrum exhibits two peaks centered at approximately 1.76~eV and 2.38~eV, which are attributed to the so-called A and B excitons,\cite{McGuire_2017, Zhu_2020_Eb, Acharya2022, refractive-index}
followed by a pronounced absorption onset at higher photon energies near 3.15~eV. 
A detailed analysis of these absorption features is presented below.

Comparison of the excitation-energy-dependent phonon intensities with the PLE and Abs spectra partially accounts for the observed variations in Raman intensity.
For all excitation energies except 3.06~eV, the evolution of the Raman peak intensities qualitatively follows the shape of the PLE spectrum, suggesting the coupling between the excited states responsible for PL emission and phonons via exciton–phonon interactions.
In contrast, the pronounced enhancement of the Raman signal under 3.06~eV excitation does not coincide with any prominent feature in either the PLE or Abs spectra, indicating that it is not of electronic-resonant origin.

Instead, this enhancement points to optical interference effects as the dominant mechanism.
To test this hypothesis, we simulated the excitation-energy dependence of the $\textrm{E}^{2}_{\textrm{g}}$, $\textrm{A}^{1/2}_{\textrm{g}}$, and $\textrm{A}^{3/4}_{\textrm{g}}$ intensities using the transfer matrix method (TMM)\cite{Velson2020} and the complex refractive index of a CrCl$_\textrm{3}$ crystal with a thickness of approximately 92~$\mu$m (see the Methods section of the SI for details).
The calculations were performed using the complex refractive index determined at room temperature ($T$=300~K).\cite{refractive-index}
As shown in Fig.~\ref{fig:raman-interference}(c), the simulated $\textrm{A}^{3/4}_{\textrm{g}}$ intensity exhibits a strong enhancement for excitation energies above approximately 2.85~eV, while reaching a minimum near 2.35~eV.
Simulations for the $\textrm{E}^{2}_{\textrm{g}}$ and $\textrm{A}^{1/2}_{\textrm{g}}$ modes are presented in Section S3 of the SI.
A detailed comparison between the calculated excitation-energy dependence of the $\textrm{A}^{3/4}_{\textrm{g}}$ intensity at 300~K and the experimentally extracted integrated intensity at 5~K shows excellent overall agreement.
In particular, the large enhancement of the Raman signal under 3.06~eV excitation is fully reproduced by the simulated interference effects, despite the difference between the experimental and simulation temperatures.
Taken together, the combined effects of resonant Raman scattering and optical interference within the CrCl$_\textrm{3}$/SiO$_\textrm{2}$/Si dielectric stack provide a comprehensive description of the excitation-energy dependence of the RS intensity.

A similar analysis of excitation-energy effects for an exfoliated CrCl$_\textrm{3}$ flake with a thickness of approximately 227~nm, together with corresponding TMM simulations, is presented in Section S4 of the SI.

The combined PL and Abs measurements enable the determination of the electronic band gap ($E_g$) of CrCl$_\textrm{3}$ as well as the exciton binding energies ($E_b$), as shown in Fig.~\ref{fig:raman-interference}(b).
The exciton binding energy is evaluated using the relation $E_b = E_g - E_{ex}$, where $E_{ex}$ denotes the excitonic peak energy.
The band gap $E_g$ is extracted using the Tauc method~\cite{Makula_2018}, expressed as $(\alpha h \nu)^{\gamma} = A(h \nu - E_g)$, where $\alpha$ is the absorption coefficient, $A$ is a proportionality constant, and $\gamma$ characterizes the nature of the electronic transition ($\gamma = 2$ for a direct band gap).
From this analysis, the electronic band gap of CrCl$_\textrm{3}$ is estimated to be $\sim$3.2~eV.
The found $E_b$ energies of the A and B excitons from the absorption measurements are approximately 1.45~eV and 0.82~eV, respectively, in good agreement with previously reported values.~\cite{refractive-index}
The PL emission peak at 1.47~eV originates from the same excitonic state as the absorption feature at 1.76~eV (A exciton) and is redshifted due to the $d-p$ hybridization between Cr and ligand states. \cite{Acharya2022, Seyler2018}
Accordingly, the binding energy associated with the A-exciton emission is estimated to be $\sim$1.73~eV.


\begin{figure*}[!th]
		\subfloat{}%
		\centering
		\includegraphics[width=1 \linewidth]{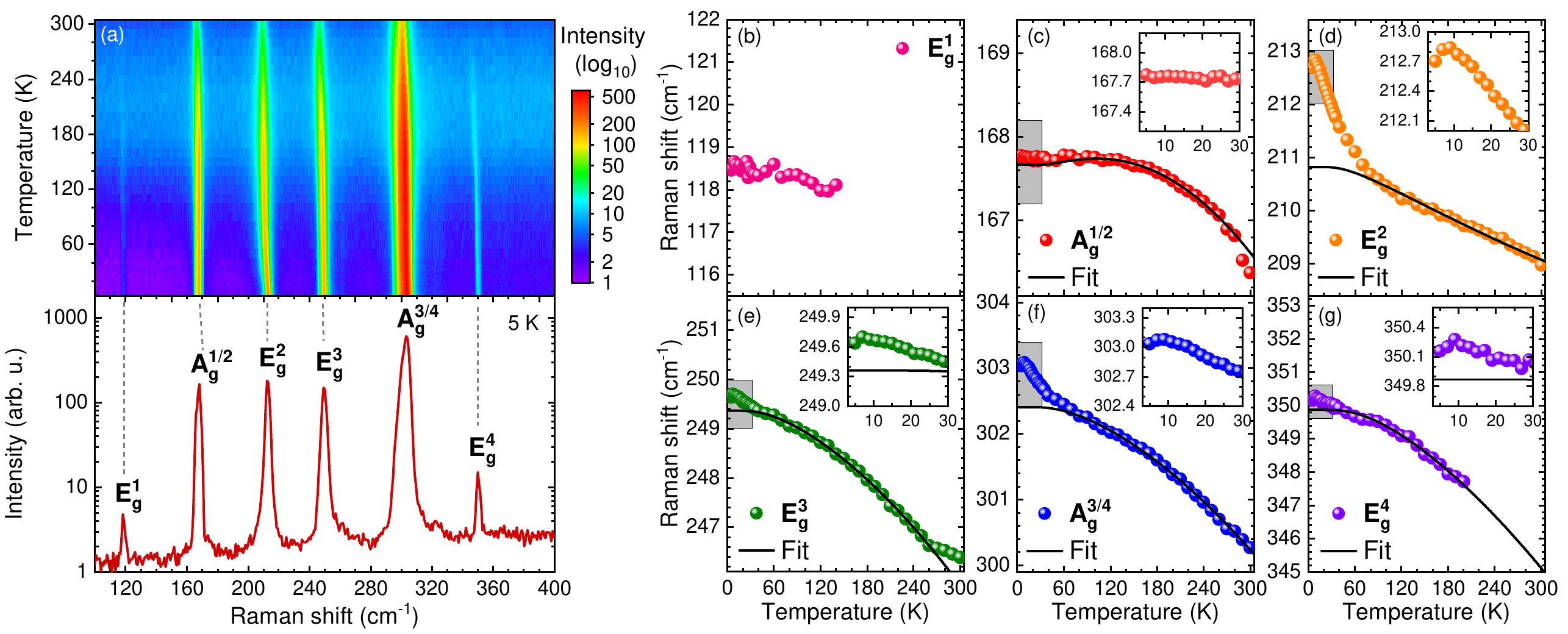}
    	\caption{Temperature evolution of the Raman modes. 
        (a) The top panel shows a false-color map of the Raman spectra of a CrCl$_\textrm{3}$ crystal, while the bottom panel presents the Raman spectrum measured at $T$=5~K. 
        The intensities are plotted on a logarithmic scale to better visualize the temperature dependence of the Raman shifts for all observed modes.
        (b)–(g) Temperature dependence of the Raman shift for all observed phonon modes. Solid lines represent fits using Eq.~\ref{eq;balkanski}. 
        The insets show the low-temperature range of the frequency evolution of the modes, as indicated by the gray rectangular regions.}
		\label{fig:temperature}
\end{figure*}

To elucidate the lattice dynamics and the coupling between phonons and magnetic order in bulk CrCl$_\textrm{3}$, we performed temperature-dependent RS measurements under 3.06~eV excitation from 5~K to 300~K.
The overall evolution of the phonon modes is summarized in the false-color intensity map shown in the top panel of Fig.~\ref{fig:temperature}(a), with a representative RS spectrum acquired at 5~K displayed in the bottom panel.
The intensities in both panels are plotted on a logarithmic scale to enhance the visibility of weaker features.
A closer inspection of the false-color map reveals distinct temperature-dependent behaviors for different phonon modes.
The intensities of the $\textrm{E}^{1}_{\textrm{g}}$ and $\textrm{E}^{4}_{\textrm{g}}$ modes decrease markedly with increasing temperature and vanish completely from the spectra at approximately 140~K and 190~K, respectively.
In contrast, the remaining modes, $i.e.$, $\textrm{A}^{1/2}_{\textrm{g}}$, $\textrm{E}^{2}_{\textrm{g}}$, $\textrm{E}^{3}_{\textrm{g}}$, and $\textrm{A}^{3/4}_{\textrm{g}}$, remain clearly observable across the entire temperature range.
Notably, the background intensity increases at elevated temperatures and reaches a maximum near 220~K, which may be associated with a broad temperature range of coexistence between the rhombohedral and monoclinic phases, extending approximately from 140~K to 240~K.\cite{McGuire_2017, Lis_2025}

Deconvolution of the temperature-dependent RS spectra using Lorentzian functions enables a detailed analysis of the temperature evolution of the Raman shifts, linewidths (full width at half-maximum, FWHM), and intensities of all six resolved phonon modes.
The corresponding Raman shifts are shown in Fig.~\ref{fig:temperature}(b)-(g) and discussed below, while the linewidths and intensities are analyzed in Section S5 of the SI.

To assess the spin-lattice coupling, the temperature dependence of the phonon energies was fitted using the standard Balkanski model, which describes anharmonic phonon softening with increasing temperature due to phonon-phonon interactions.\cite{Balkanski1983}
The model is given by
\begin{equation}
\label{eq;balkanski}
\begin{split}
\omega_{anh}(T) = \omega_{0} + A \left(1+\frac{2}{e^{x}+1} \right)\\ + B \left(1 + \frac{3}{e^{y}-1} + \frac{3}{(e^{y}-1)^2} \right),
\end{split}
\end{equation}
where $\omega_{0}$, $A$, and $B$ are fitting parameters, $x = \frac{\hbar\omega_{0}}{2k_{B}T}$, and $y = \frac{\hbar\omega_{0}}{3k_{B}T}$.
The quantity $\omega_{0}+A+B$ corresponds to the phonon frequency extrapolated to 0~K.

The temperature evolution of the $\textrm{A}^{1/2}_{\textrm{g}}$ mode is well described by the Balkanski model, indicating that this phonon is only weakly affected by magnetic interactions and behaves predominantly as a classical anharmonic mode governed by phonon-phonon scattering.\cite{Yan2022, Wilczynski_2025}
In contrast, the remaining modes ($\textrm{E}^{2}_{\textrm{g}}$, $\textrm{E}^{3}_{\textrm{g}}$, $\textrm{A}^{3/4}_{\textrm{g}}$, and $\textrm{E}^{4}_{\textrm{g}}$) exhibit clear deviations from the standard anharmonic behavior,\cite{Balkanski1983, Taube_2015, Joshi_2016, Sarkar_2020} particularly in the low-temperature regime between 5~K and 80~K.

A detailed inspection of this temperature range reveals that, for all these modes, the Raman shift increases slightly as the temperature rises from 5~K to approximately 10~K.
This temperature range is close to the N\'{e}el temperature ($T_\mathrm{N}$=14~K) of CrCl$_\textrm{3}$, which marks the transition from the AFM phase to the paramagnetic order.\cite{Pocs_Neel_CrCl3, Bykovetz_2019} 
The magnitude of the anomalous redshift between 5~K and 80~K varies substantially among the modes.
The $\textrm{E}^{2}_{\textrm{g}}$ mode exhibits the strongest anomaly, with an energy shift of approximately 2.0~cm$^{-1}$, whereas the $\textrm{E}^{3}_{\textrm{g}}$ and $\textrm{E}^{4}_{\textrm{g}}$ modes show intermediate shifts of about 0.6~cm$^{-1}$.
The $\textrm{A}^{3/4}_{\textrm{g}}$ mode displays a comparable shift of approximately 0.7~cm$^{-1}$.
At higher temperatures (80~K-300~K), the slopes of the experimental data change, and the phonon energies recover a conventional anharmonic temperature dependence similar to that observed in both magnetic and nonmagnetic materials.\cite{Taube_2015, Joshi_2016, Sarkar_2020}

The presence of anomalous temperature-dependent behavior, not accounted for by anharmonic effects, in the majority of phonon modes up to approximately 80~K strongly suggests that magnetic ordering in CrCl$_\textrm{3}$ persists to temperatures well above the N\'{e}el temperature.
The most plausible scenario therefore involves three distinct magnetic regimes in CrCl$_\textrm{3}$.
At $T < T_\mathrm{N}$, CrCl$_\textrm{3}$ exhibits fully developed AFM order, characterized by FM ordering within individual layers and AFM coupling between adjacent layers.
Notably, ferromagnetic ordering persisting above $T_\mathrm{N}$ has already been reported for thick exfoliated CrCl$_\textrm{3}$ flakes.\cite{Kipczak2025}
In the intermediate temperature range $T_\mathrm{N} < T < 80$~K, a locally ordered magnetic state emerges, in which FM correlations persist within individual layers in a domain-like structure, while interlayer AFM order is suppressed.
Because the phonon modes examined here involve exclusively intralayer atomic vibrations within a single CrCl$_\textrm{3}$ layer, the observed anomalies indicate that local FM order within individual layers survives up to temperatures of approximately 80~K.
At higher temperatures ($T > 80$~K), CrCl$_\textrm{3}$ enters a purely paramagnetic phase.

The temperature evolution of the Raman modes provides a clear signature of the structural phase transition in bulk CrCl$_\textrm{3}$ from the low-temperature rhombohedral ($\bar{R}3$) phase to the high-temperature monoclinic ($C2/m$) phase, with phase coexistence between approximately 140~K and 240~K.\cite{McGuire_2017, Lis_2025}
As shown in Fig.~\ref{fig:temperature}(a), the RS spectra undergo a pronounced transformation upon warming.
The coexistence regime is particularly evident in the $\textrm{E}^{3}_{\textrm{g}}$ mode (Fig.~\ref{fig:temperature}(e)), which exhibits a change in slope near 250~K, indicating completion of the transition to the monoclinic phase.
Low-intensity modes such as $\textrm{E}^{1}_{\textrm{g}}$ and $\textrm{E}^{4}_{\textrm{g}}$ become unobservable at elevated temperatures due to weak Raman cross sections and thermal broadening, although they are expected to persist, consistent with previous reports\cite{Glamazda_2017} and phonon calculations for the monoclinic phase.\cite{Kazim_2020}
In addition, the $\textrm{E}^{2}_{\textrm{g}}$, $\textrm{E}^{3}_{\textrm{g}}$, and $\textrm{A}^{3/4}_{\textrm{g}}$ modes exhibit strong intensity suppression above $\sim$160~K, marking the onset of structural instability in the rhombohedral lattice (see Section S5 of the SI).
Overall, the Raman data confirm that the rhombohedral ($\bar{R}3$) phase is stable up to $\sim$140~K, followed by a broad transition to monoclinic ($C2/m$) symmetry that is fully established above 240~K.\cite{McGuire_2017, Lis_2025}

The corresponding analysis of the temperature dependence of the RS spectra of the exfoliated CrCl$_\textrm{3}$ flake, presented in Section~S6 of the SI, demonstrates behavior similar to that discussed above for the bulk CrCl$_\textrm{3}$ crystal.

\begin{table}[t]
\caption{Spin-phonon coupling constant ($\lambda$) values of Raman modes}
\label{table:2}
\begin{tabular}{ c || c | c | c | c }
 & $\textrm{E}^{2}_\textrm{g}$ & $\textrm{E}^{3}_\textrm{g}$ & $\textrm{A}^{3/4}_\textrm{g}$ & $\textrm{E}^{4}_\textrm{g}$ \\
 \hline \hline
 $\lambda$  &   0.94~cm$^{-1}$  &  0.28~cm$^{-1}$  & 0.35~cm$^{-1}$ & 0.28~cm$^{-1}$ \\
\end{tabular}
\end{table}

The strength of the spin-phonon interaction at 5~K was quantified by extracting the spin–phonon coupling coefficients ($\lambda$) using the commonly employed relation:\cite{Yin2021, Kozlenko2021, Kipczak2024, Wei_2025-CrSBr-s-ph}
\begin{equation}
\omega(T) \simeq \omega_{\mathrm{anh}}(T) + \lambda \langle s_i \cdot s_j \rangle ,
\end{equation}
where $\omega(T)$ is the measured phonon frequency, $\omega_{\mathrm{anh}}(T)$ is the anharmonic phonon frequency in the absence of spin-phonon coupling, $\lambda$ is the spin-phonon coupling constant, and $\langle s_i \cdot s_j \rangle$ denotes the nearest-neighbor spin-spin correlation function.\cite{Yin2021, Kozlenko2021, Kipczak2024, Wei_2025-CrSBr-s-ph, Lili2023}.
For Cr$^{3+}$ ions with spin $S = 3/2$, $\langle s_i \cdot s_j \rangle = 9/4$, consistent with the magnetic moment of approximately $3\mu_B$.\cite{Grodzicki_2010, McGuire_2017, Pocs_Neel_CrCl3, Lu2020, Bedoya-Pinto_2021, Buccoliero_2025}
The extracted values of the spin–phonon coupling constants are summarized in Table~\ref{table:2}.
The strongest coupling is observed for the $\textrm{E}^{2}_{\textrm{g}}$ mode, with $\lambda \approx 0.94$~cm$^{-1}$.
The $\textrm{A}^{3/4}_{\textrm{g}}$, $\textrm{E}^{3}_{\textrm{g}}$, and $\textrm{E}^{4}_{\textrm{g}}$ modes exhibit smaller yet still significant coupling strengths of approximately 0.35~cm$^{-1}$, 0.28~cm$^{-1}$, and 0.28~cm$^{-1}$, respectively.

To contextualize these findings, we compare our results with those reported in the literature for the other magnetic materials. 
The coupling strengths determined for the modes A$^{3/4}_\textrm{3}$, E$^{3}_\textrm{g}$ and E$^{4}_\textrm{g}$ are comparable to the values observed in CrBr$_\textrm{3}$, where $\lambda$ for the E$_\textrm{g}$ mode is reported as 0.27 cm$^{-1}$,\cite{Kozlenko2021} 0.43 cm$^{-1}$,\cite{Kipczak2024} and up to 0.51 cm$^{-1}$, \cite{Yin2021} as well as in CrSBr ($\lambda$~=~0.29 cm$^{-1}$ fo the A$_\textrm{g}$ mode).\cite{Wei_2025-CrSBr-s-ph}
Remarkably, the strong coupling identified for the E$^{2}_\textrm{g}$ mode approaches the substantial values found in heavy-element tellurides, such as Cr$_\textrm{2}$Ge$_\textrm{2}$Te$_\textrm{6}$, where $\lambda$ reaches 0.82 cm$^{-1}$ \cite{krasucki2025} and 1.2 cm$^{-1}$ for the E$_\textrm{g}$ \cite{Tian2016} mode and Fe$_\textrm{2}$GeTe$_\textrm{2}$ - $\lambda$~=~1.3 cm$^{-1}$ for the A$_\textrm{g}$ mode.\cite{Du2019-FeGeTe}

In summary, we have systematically investigated the vibrational properties of bulk and thick-flake CrCl$_\textrm{3}$ using polarization-resolved RS spectroscopy in combination with PL, PLE, and Abs measurements. 
Using symmetry analysis and first-principles phonon calculations, we achieve an unambiguous experimental assignment of all eight Raman-active phonon modes, comprising four $\textrm{A}_\textrm{g}$ and four $\textrm{E}_\textrm{g}$ modes. Excitation-energy-dependent measurements demonstrate that the pronounced enhancement of selected phonon modes is governed primarily by optical interference effects rather than by resonant Raman scattering.
Temperature-dependent Raman spectroscopy reveals strong signatures of spin–phonon coupling, indicating the persistence of local magnetic correlations well above the antiferromagnetic ordering temperature and delineating a transition from a fully antiferromagnetic phase through an intermediate domain-like ferromagnetic regime to a paramagnetic state. 
In addition, clear Raman signatures of a rhombohedral-to-monoclinic structural phase transition are observed. 
These results highlight the intricate interplay between lattice, electronic, and magnetic degrees of freedom in CrCl$_\textrm{3}$ and establish Raman spectroscopy as a powerful probe of spin-lattice coupling in layered magnetic materials.

\section*{Data availability}
The data that support the findings of this work are available from the corresponding authors upon reasonable request.

\section*{Acknowledgments}
The work was supported by the National Science Centre, Poland (grant no. 2020/37/B/ST3/02311, 2023/48/C/ST3/00309).
T.W. gratefully acknowledges Poland’s high-performance Infrastructure PLGrid ACC Cyfronet AGH for providing computer facilities and support within computational Grant No. PLG/2025/018073.
Z.S. was supported by project LUAUS25268 from Ministry of Education Youth and Sports (MEYS) and by the project Advanced Functional Nanorobots (reg. no. CZ.02.1.01/0.0/0.0/15\_003/0000444 financed by the EFRR). 
Z.S. acknowledges the assistance provided by the Advanced Multiscale Materials for Key Enabling Technologies project, supported by the Ministry of Education, Youth, and Sports of the Czech Republic (project no. CZ.02.01.01/00/22\_008/0004558), co-funded by the European Union.
M.K. acknowledges the Ministry of Education (Singapore) through the Research Centre of Excellence program (grant EDUN C-33-18-279-V12, I-FIM) and under its Academic Research Fund Tier 2 (MOE-T2EP50122-0012), and the Air Force Office of Scientific Research and the Office of Naval Research Global under award number FA8655-21-1-7026. 

\section*{Author Contributions}
M.R.M. initiated and supervised the project. 
Ł.K. prepared the CrCl$_\textrm{3}$ samples and performed the RS and PL measurements.
Ł.K. and J.P. performed the AFM imaging.
T.W. calculated the phonon dispersion using density functional theory (DFT). 
Ł.K. and I.A. carried out the PLE experiment. 
C.K.M., J.I., and A.W. carried out the absorption measurements. 
P.O. and Ł.K. simulated the interference effects. 
M.K. (Prague) and Z.S. grew the CrCl$_\textrm{3}$ crystals. 
Ł.K. analyzed the experimental data with contributions from A.B., M.K. (Singapore), and M.R.M. 
Ł.K and M.R.M. wrote the manuscript with input from all co-authors.

\bibliographystyle{apsrev4-2}
\bibliography{biblio}

\newpage
\onecolumngrid

\renewcommand{\thefigure}{S\arabic{figure}}

\makeatletter
\def\@hangfrom@section#1#2#3{\@hangfrom{#1#2}#3}
\def\@hangfroms@section#1#2{#1#2}
\makeatother

\begin{center}
	{\large{ {\bf Supporting Information: \\ Strong Spin-Lattice Interaction in Layered Antiferromagnetic CrCl$_\textrm{3}$}}}
	\vskip0.5\baselineskip{\L{}ucja Kipczak\orcidlink{0000-0003-1266-0201},{$^{1}$} Tomasz Wo\'zniak,{$^{1,2}$} Chinmay K. Mohanty,{$^{1}$} Igor Antoniazzi,{$^{1}$} \\ Jakub~Iwa\'nski\orcidlink{0000-0003-2395-4010},{$^{1}$} Przemys\l{}aw~Oliwa\orcidlink{0000-0003-1255-5997}, {$^{1}$} Jan Paw\l{}owski,{$^{1}$} Meganathan Kalaiarasan,{$^{3}$} Zden\v{e}k~Sofer,{$^{3}$} \\ Andrzej~Wysmo\l{}ek\orcidlink{0000-0002-8302-2189},{$^{1}$} Adam~Babi\'nski\orcidlink{0000-0002-5591-4825},{$^{1}$} Maciej Koperski\orcidlink{0000-0002-8301-914X},{$^{4,5}$} and Maciej R. Molas\orcidlink{0000-0002-5516-9415},{$^{1}$}
	}
	
	\vskip0.5\baselineskip{\textit{$^{1}$ Faculty of Physics, University of Warsaw, Pasteura 5, 02-093 Warsaw, Poland \\$^{2}$ Faculty of Fundamental Problems of Technology, Wroc\l{}aw Univwersity of Science and Technology, Wyb. Wyspia{\'n}skiego 27, 50-370 Wroc\l{}aw, Poland \\$^{3}$ Department of Inorganic Chemistry, University of Chemistry and Technology, Technická 5, 160 00 Praha 6-Dejvice, Prague, Czech Republic \\$^{4}$ Institute for Functional Intelligent Materials, National University of Singapore, 9 Engineering Dr 1, 117544, Singapore \\$^{5}$ Department of Materials Science and Engineering, National University of Singapore, 9 Engineering Dr 1, 117575, Singapore}}
\end{center}

This Supporting Information provides: \ref{sec:methods} - Details of the crystal synthesis, samples fabrication, experimental setups, transfer matrix method simulations, and first principles calculations. \ref{sec:S1} - Raman scattering spectra of a commercial CrCl$_\textrm{3}$ crystal. \ref{sec:S2} - Angle-resolved polarization of phonon modes in CrCl$_{3}$ crystal. \ref{sec:S3} - Transfer matrix method simulations of phonon mode intensities in CrCl$_\textrm{3}$. \ref{sec:S4} - Angle-resolved and excitation-dependent Raman spectra of an exfoliated CrCl$_\textrm{3}$ flake. \ref{sec:S5} - Temperature evolution of the Raman mode intensities and linewidths in the  CrCl$_\textrm{3}$ crystal. \ref{sec:S6} - Temperature-dependent Raman spectra of an exfoliated CrCl$_\textrm{3}$ flake.


\renewcommand{\thesection}{Methods}
\setcounter{figure}{0}
\renewcommand{\theHfigure}{S\arabic{figure}} 
\section{\label{sec:methods}}
\subsection*{Crystal synthesis}
CrCl$_\textrm{3}$ crystals were prepared from polycrystalline material using the chemical vapor transport method in quartz ampoule. 
30 g of CrCl$_\textrm{3}$ (99.9\%, Strem, USA) were placed in quartz ampoule (50$\times$250 mm) and degassed under high vacuum (<1$\times$10$^{-3}$ Pa, oil diffusion pump with LN$_\textrm{2}$ cold trap) at 100$^\circ$C for 2 hours and subsequently melt sealed using oxygen-hydrogen welding torch. 
The ampule was placed in a two zone horizontal furnace for crystal growth. 
The growth zone was first heated at 1000$^\circ$C and the source zone at 700$^\circ$C for 2 days. 
Subsequently, the thermal gradient was reversed and the source zone was heated at 900$^\circ$C and the growth zone at 800$^\circ$C for 15 days. 
Finally, the ampule was cooled to room temperature and opened in an argon filled glovebox.

\subsection*{Samples fabrication}

\begin{figure*}[h]
		\subfloat{}%
		\centering
		\includegraphics[width=0.8 \linewidth]{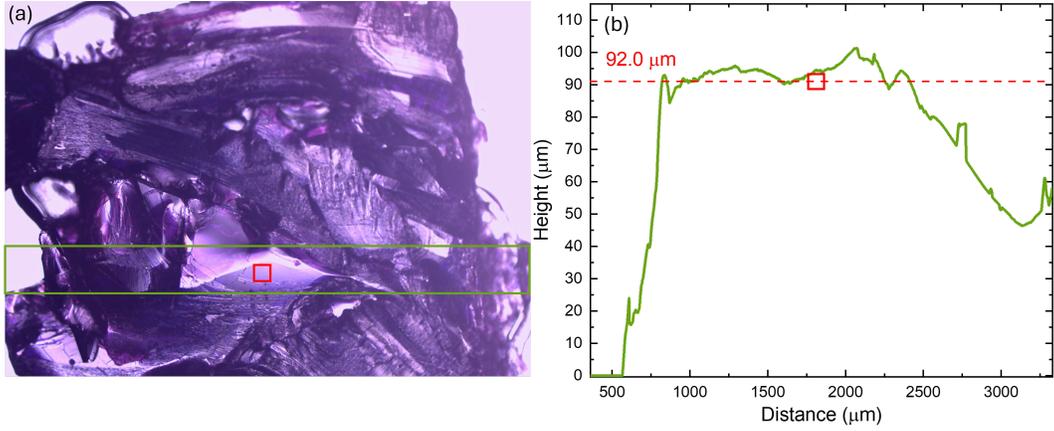}
    	\caption{(a) Optical image of the investigated CrCl$_\textrm{3}$ crystal. The green rectangle indicates the scanning area profiled by the stylus (center of rectangle), while the red square marks the spot selected for Raman measurements.
        (b) Topographical profile corresponding to the region marked in (a). The green curve shows the height variation across the sample. The dashed red line indicates the thickness at the Raman measurement spot (marked by the red square), which is 92.0~$\mu$m.
        Note that the distance axis has been offset to align the scan range with the optical image's field of view.}
		\label{fig:crystal-profile}
\end{figure*}

For Raman scattering (RS), photoluminescence (PL), and photoluminescence excitation (PLE) measurements, a bulk CrCl$_\textrm{3}$ crystal was mechanically placed on a silicon substrate coated with a 90~nm-thick SiO$_\textrm{2}$ layer. 
An optical micrograph of the sample is shown in Fig.~\ref{fig:crystal-profile}(a).

The crystal thickness at the Raman measurement position, indicated by the green rectangle in Fig.~\ref{fig:crystal-profile}(a), was determined by stylus profilometry. 
The height profile was acquired using a Veeco Dektak~6M profilometer equipped with a stylus 10~$\mu$m $\times$ 2~$\mu$m ASP, over a scan area of 16~mm$^2$ as shown in Fig.~\ref{fig:crystal-profile}(b). 
The thickness at this location was approximately 92.0~$\mu$m. The maximum height observed across the full scan range reached approximately 140~$\mu$m, which lies outside the region displayed in the optical image.

For absorption measurements, the CrCl$_{3}$ crystal was mounted on a copper plate containing a 1~mm-diameter aperture.

\begin{figure*}[h]
		\subfloat{}%
		\centering
		\includegraphics[width=0.8 \linewidth]{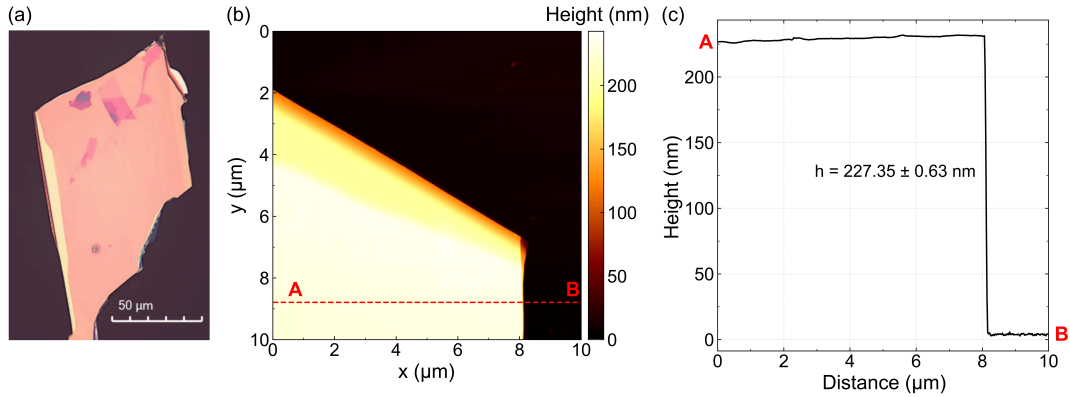}
    	\caption{(a) Optical image of the investigated CrCl$_\textrm{3}$ flake. The red square indicates the probed area. (b) Atomic force microscopy (AFM) topography of the investigated flake. The red line (A--B) marks the location of the cross-section used for thickness determination. (c) Height profile extracted along the line (A-B) revealing a flake thickness of 227.35$\pm$0.63 nm. }
		\label{fig:afm}
\end{figure*}

Thick exfoliated CrCl$_\textrm{3}$ flakes were prepared from bulk crystals using a polydimethylsiloxane (PDMS)-based exfoliation method and subsequently transferred onto SiO$_\textrm{2}$/Si substrates with a 285~nm-thick SiO$_\textrm{2}$ layer via an all-dry deterministic stamping technique, which avoids glue residues.\cite{Castellanos-Gomez_2014}
The thickness of the investigated CrCl$_\textrm{3}$ flake was determined by atomic force microscopy (AFM) using a Bruker Dimension Icon system equipped with a NanoScope~VI controller. 
AFM measurements were performed in PeakForce Tapping mode under ambient conditions using a silicon probe (RFESP-75 model).
The flake thickness was found to be approximately 227~nm, as shown in Fig.~\ref{fig:afm}.

\newpage
\subsection*{Raman scattering} 
RS spectra were excited using diode lasers with wavelengths of $\lambda = 405$~nm (3.06~eV), $\lambda = 515$~nm (2.41~eV), and $\lambda = 561$~nm (2.21~eV), as well as a He-Ne laser ($\lambda = 633$~nm, 1.96~eV) and an Ar-ion laser ($\lambda = 488$~nm, 2.54~eV).
Samples were mounted on the cold finger of a continuous-flow cryostat, enabling measurements over a temperature range from $\sim$5~K to room temperature (300~K).
The excitation beam was focused using a 50$\times$ long-working-distance objective with a numerical aperture (NA) of 0.55, yielding a spot size of $\sim$1~$\mu$m in diameter.
The scattered light was collected by the same objective in a backscattering geometry, dispersed by a 0.75~m spectrometer equipped with an 1800~grooves/mm grating, and detected with a liquid-nitrogen-cooled charge-coupled device (CCD).
Laser lines were suppressed prior to the spectrometer using long-pass filters (488~nm, 515~nm, 561~nm, and 633~nm) or Bragg-grating notch filters (405~nm).
Polarization-resolved RS measurements were performed using a motorized half-wave plate and a fixed linear polarizer in the detection path.

\subsection*{Photoluminescence excitation.} 
PLE measurements were performed using a broadband supercontinuum light source spectrally filtered by a monochromator, producing a monochromatic excitation beam with a linewidth of approximately 2~nm.
The excitation power at the sample was stabilized using an electrically controlled liquid crystal followed by a linear polarizer, operating in a feedback loop with a photodiode monitoring the optical power entering the probe.
The PLE spectra were recorded using the same experimental setup as for the RS measurements, with the spectrometer grating replaced by a 300~grooves/mm grating.

\subsection*{Absorbance}
Abs spectra of the CrCl$_\textrm{3}$ crystal were recorded over the 350--850 nm spectral range using a Cary 5000 UV--Vis--NIR spectrophotometer operated in double-beam mode. 
Low-temperature measurements were performed using an Oxford Instruments closed-cycle helium cryostat equipped with an ITC502 temperature controller.
To eliminate contributions from the sample mount and the finite aperture size, Abs spectra were acquired for both the crystal and an empty aperture of identical dimensions. 
The reference spectrum was subsequently subtracted from the sample spectrum, yielding the absorbance signal originating exclusively from the crystal.

\subsection*{Transfer-matrix method}
To investigate the impact of layer thickness and optical interference on the measurable intensity of Raman modes, a rigorous model based on the TMM was employed. 
This approach accounts for multiple reflections of both the incident laser field and the scattered Raman signal at all interfaces within the multilayer structure, as well as for Raman signal generation throughout the volume of the active material. 
The calculations were performed following the formalism described by Van Velson et al.
\cite{Velson2020}.

The considered system consists of layers characterized by complex refractive indices $\tilde n_{j} = n_{j}+i\kappa_{j}$, where $j$ denotes the layer index (air, CrCl$_{3}$, SiO$_{2}$/Si substrate). 
Assuming normal incidence of light, the relationship between the electric field amplitudes $E^{+}$ (forward propagating wave) and $E^{-}$ (backward propagating wave) at the layer boundaries is described by the interference matrix $I_{jk}$ and the propagation matrix $L_{j}$:
\begin{equation}
    I_{jk} = \frac{1}{t_{jk}} \left(\begin{matrix}
        1 & r_{jk} \\
        r_{jk} & 1
    \end{matrix} \right), L_{j}=\left(\begin{matrix}
        e^{-i\beta{j}d{j}} & 0\\
        0 & e^{i\beta_{j}d{j}}
    \end{matrix} \right),
\end{equation}
where $r_{jk}$ and $t_{jk}$ are the Fresnel reflection and transmission coefficients, $d_{j}$ is the layer thickness, and $\beta_{j} = \frac{2\pi}{\lambda}\tilde n_{j}$ represents the propagation constant.

The calculation of the total Raman signal intensity ($I_{Raman}$) was divided into two stages.
First, excitation field distribution. 
The local electric field amplitude of the incident laser, $E_{exc.}(x)$, was determined as a function of depth $x$ within the active layer (CrCl$_\textrm{3}$). 
The local RS generation intensity at depth $x$ is assumed to be proportional to the square of the electric field amplitude, $|E_{exc.}(x)|^{2}$.
Second, Raman signal propagation.
The Raman signal generated at depth $x$ at the Raman wavelength $\lambda_{Raman}$ was treated as a plane-wave source emitted in two directions: forward (into the substrate) and backward (toward the detector). 
The algorithm rigorously accounts for the discontinuity of the electric field at the generation point $x$.
The total Raman signal intensity reaching the detector results from the interference of waves generated at all depths within the active layer. 
In the simulations, an incoherent summation of contributions from different depths was assumed, which is the standard approach for layers with thicknesses significantly exceeding the coherence length of the Raman process. 
The final intensity was calculated as an integral over the active layer thickness $d_{active}$ using formula:
\begin{equation}
\begin{split}
    I_{Raman} \propto \int_{0}^{d_{active}}\mid E_{exc.}(x,\lambda_{exc.})\mid^{2}\cdot [ \mid E_{out, back}(x,\lambda_{Raman})\mid^{2} + \mid E_{out,forward}(x, \lambda_{Raman)}\mid^{2}] dx, 
\end{split}
\end{equation}
where $E_{out}$ represents the efficiency of coupling the Raman signal from depth $x$ to the outside of the system.
Numerical integration was performed using adaptive quadrature methods implemented in Python. 
The simulations accounted for the dispersion of all constituent materials by interpolating the wavelength-dependent complex refractive index, $\tilde{n}(\lambda) = n(\lambda) + i\kappa(\lambda)$, where $n(\lambda)$ and $\kappa(\lambda)$ correspond to the refractive index and extinction coefficient, respectively, evaluated at both the excitation wavelength and the Raman-scattered wavelengths. 
The complex refractive indices were taken from the literature for CrCl$_\textrm{3}$ \cite{refractive-index}, SiO$_\textrm{2}$\cite{refractive_SiO2}, and Si\cite{refractive_Si}.

\subsection*{DFT calculations} 
The first-principles calculations were performed within the density functional theory (DFT) in the Vienna ab-initio simulation package (VASP)~\cite{Kresse1993, Kresse1994, Kresse1996, Kresse1996v2}.
The projector augment wave (PAW) potentials and general gradient approximation (GGA) of Pedew-Burke-Ernzerhof (PBE)~\cite{Kresse1999} with D3 van der Waals correction~\cite{Grimme2010} were used.
Spin-orbit coupling was not included due to its insignificant impact on geometrical parameters and phonon dispersion in this compound, as shown by Li $et$ $al.$ in Ref.~\cite{Li2025}.
The primitive cell vectors and atomic positions have been optimized until forces on atoms were lower than 10$^{-5}$~eV/Å and stress tensor components were lower than 0.1~kbar.
A plane wave basis set cutoff of 500~eV and a $\Gamma$-centered k-mesh 6$\times$6$\times$6 were sufficient to converge the lattice constants with precision of 0.001~\AA.
An energy tolerance of 10$^{-8}$~eV was used to converge the charge density in all calculations.
The phonon dispersion was calculated in a 2$\times$2$\times$2 supercell using the finite displacement method as implemented in Phonopy package~\cite{Togo2023,Togo2023first}.
All the calculations were performed at temperature of 0~K.


\newpage
\renewcommand{\thesection}{S\arabic{section}}
\setcounter{section}{0}
\renewcommand{\theHfigure}{S\arabic{figure}} 
\setcounter{secnumdepth}{3}

\section{Raman scattering spectra of a commercial CrCl\texorpdfstring{$_3$}{3} crystal \label{sec:S1}}

\begin{figure*}[h]
		\subfloat{}%
		\centering
		\includegraphics[width=0.65 \linewidth]{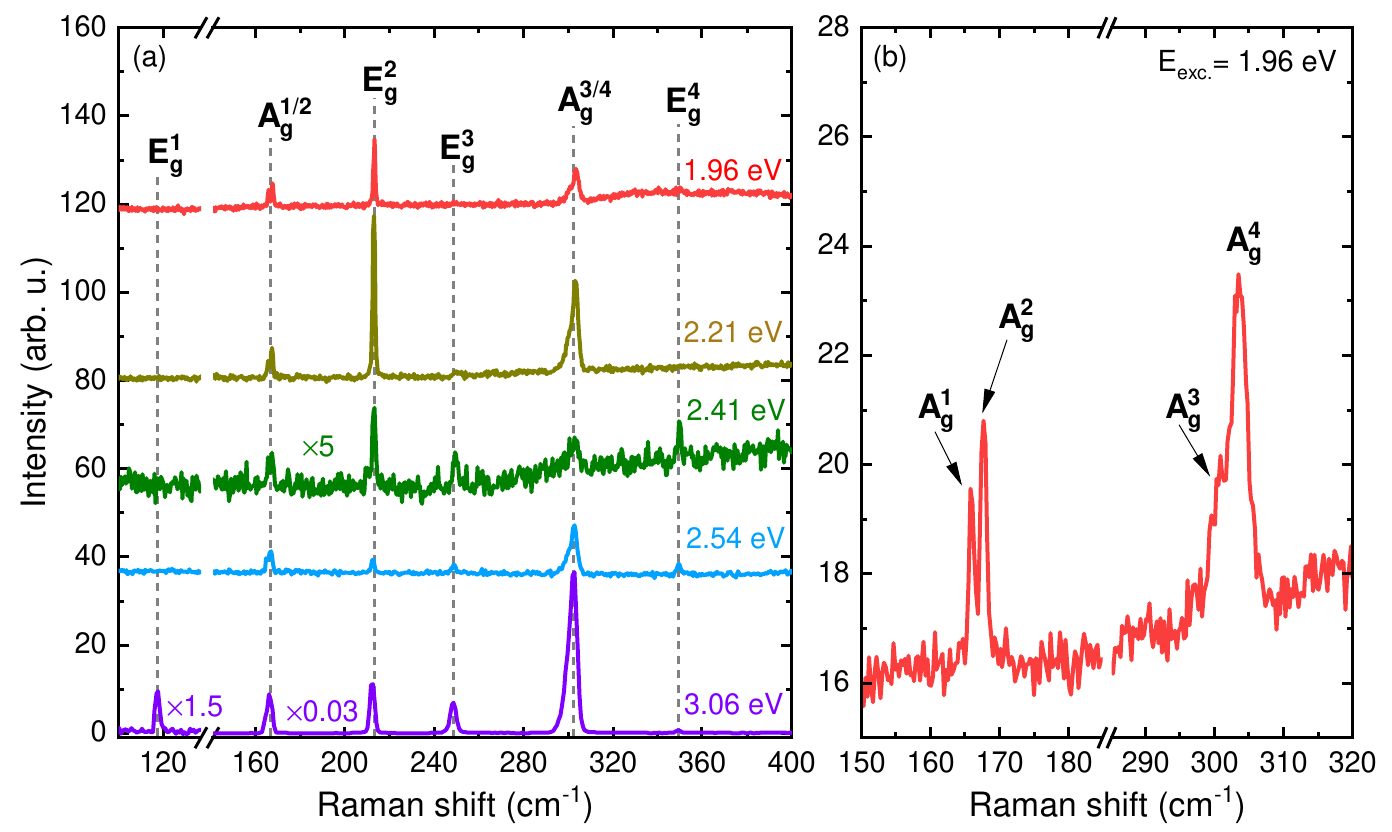}
    	\caption{      
        Resonant Raman scattering measurements of a commercial CrCl$_\textrm{3}$ crystal.
        (a) Raman spectra of the CrCl$_\textrm{3}$ crystal acquired at $T=5$~K using excitation energies of 1.96~eV, 2.21~eV, 2.41~eV, 2.54~eV, and 3.06~eV with an excitation power of 75~$\mu$W. 
        The spectra are vertically shifted for clarity.
        (b) Low-temperature ($T=5$~K) Raman spectrum of the same CrCl$_\textrm{3}$ crystal measured using 1.96~eV excitation, highlighting the doublet structure of two A$_\textrm{g}$ modes.}
		\label{fig:raman-commercial}
\end{figure*}

To validate the influence of excitation energy on the Raman spectra across different samples, we performed a comparative study using a commercial CrCl$_\textrm{3}$ crystal purchased from HQ Graphene. 
Figure~\ref{fig:raman-commercial}(a) shows low-temperature Raman spectra acquired using excitation energies of 3.06~eV, 2.54~eV, 2.41~eV, 2.21~eV, and 1.96~eV. 
Overall, the relative intensities of the Raman modes are comparable to those observed in the as-synthesized crystal discussed in the main text.
Consistent with previous observations, the spectrum acquired with the 3.06~eV excitation was scaled by a factor of 0.03 for clarity. 
In addition, the spectral region between 100--130~cm$^{-1}$ was multiplied by 1.5 to enhance the visibility of the weak E$^{1}_\textrm{g}$ mode.
Furthermore, the spectrum measured using the 2.41~eV excitation was scaled by a factor of 5 to facilitate comparison.
Despite this enhancement, the corresponding peaks remain only slightly above the background noise level.
Nevertheless, the corresponding peaks remain only slightly above the background noise level.

The intensity dependence on excitation energy follows a trend analogous to that shown in Fig.~2(a) of the main text.
Moreover, the doublet structure of the A$_\textrm{g}$ modes near 168~cm$^{-1}$ and 303~cm$^{-1}$ is clearly resolved for the 1.96~eV excitation, as shown in Fig.~\ref{fig:raman-commercial}(b).

Taken together, these results confirm the presence of all eight Raman-active modes predicted by the phonon dispersion calculations (Fig.~1(b) in the main text), which are consistently observed in CrCl$_\textrm{3}$ crystals obtained from two independent sources.

\newpage
\section{Angle-resolved polarization of phonon modes in CrCl\texorpdfstring{$_3$}{3} crystal\label{sec:S2}}

\begin{figure*}[h]
		\subfloat{}%
		\centering
		\includegraphics[width=1 \linewidth]{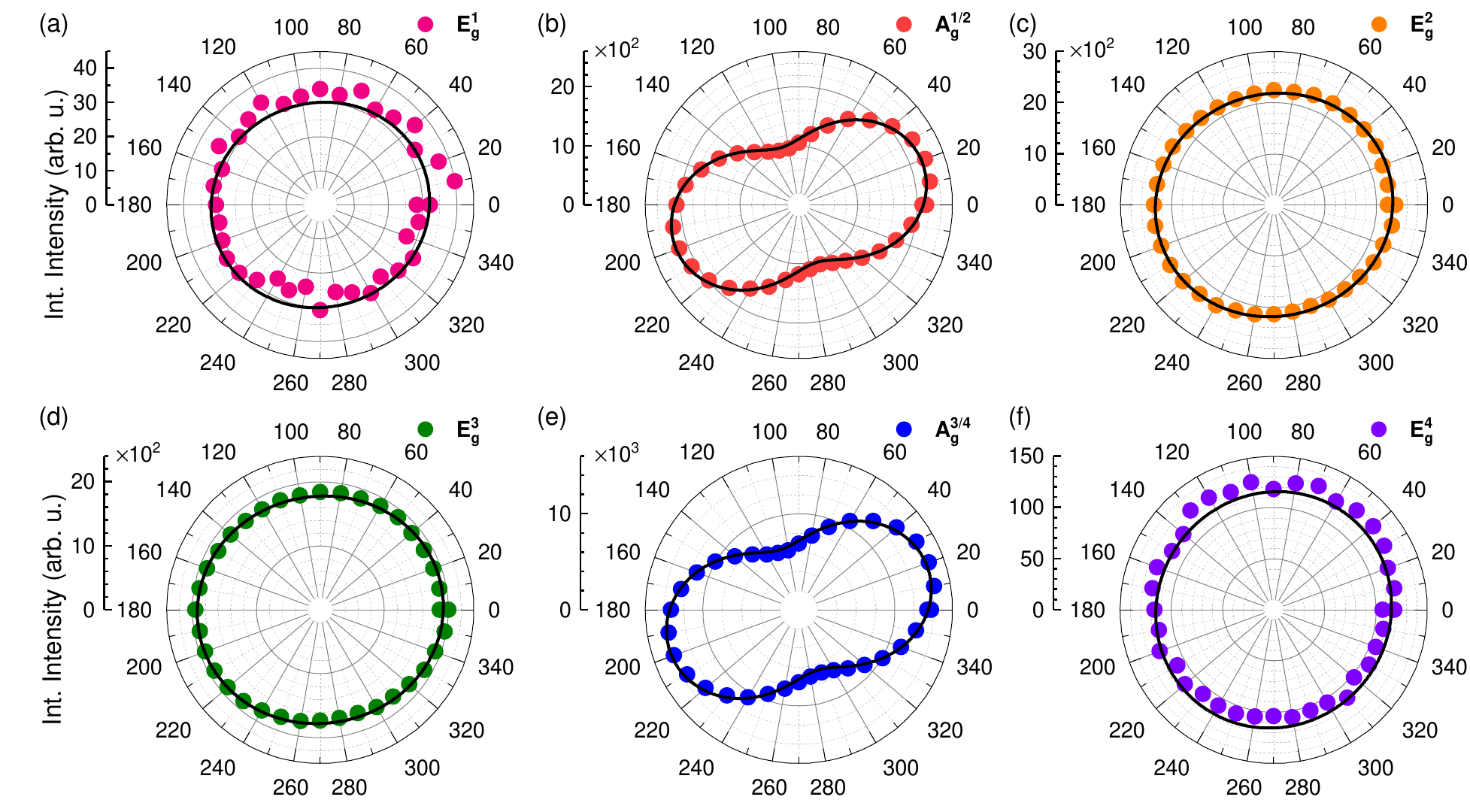}
    	\caption{Polarization dependence of the six Raman-active modes observed in the Raman spectra of a CrCl$_\textrm{3}$ crystal. Measurements were performed using a 3.06~eV laser at low temperature ($T$=5~K) with an excitation power of approximately 75~$\mu$W. The modes shown in panels (a), (c), (d), and (f) correspond to the E$_\textrm{g}$ symmetry, while those in panels (b) and (e) are of A$_\textrm{g}$ symmetry. The black curves represent fits to Eq.~(\ref{eq:pol}).}
		\label{fig:polarization}
\end{figure*}

To determine the symmetry of the Raman-active modes in CrCl$_\textrm{3}$, we measured their linear polarization dependence at low temperature ($T=5$~K). 
The intensities of the Raman modes presented in Figs.~\ref{fig:polarization}(a), (c), (d), and (f) remain unchanged as a function of polarization angle. 
Therefore, the modes were assigned to the E$_\textrm{g}$ symmetry and are associated with in-plane atomic vibrations within the crystal lattice.
In contrast, the intensities of the modes shown in Figs.~\ref{fig:polarization}(b) and (e) exhibit a clear dependence on the polarization angle. 
For these modes, the experimental data was fitted using Eq.~(\ref{eq:pol}),
\begin{equation}
I(\theta)=I_0 + A \cos^2(\theta-\phi),
\label{eq:pol}
\end{equation}
where $I_0$ denotes the background intensity, $A$ is the amplitude of the intensity modulation, and $\phi$ is a phase offset.
These modes are therefore assigned to the A$_\textrm{g}$ symmetry and correspond to out-of-plane atomic vibrations.

In particular, the intensity of the A$_\textrm{g}$ modes does not decrease to zero near a polarization angle of approximately 110$^\circ$. 
This behavior is likely due to the resonant excitation effects, which was previously reported in the literature, for thin layers of TMDs.\cite{Lee2018-tmd-pol, Tan2021}
Because the polarization-dependent measurements were performed using a 3.06~eV excitation laser, for which the spectral resolution is lower than that obtained with the 1.96~eV excitation, the doublet structure of the A$_\textrm{g}$ modes could not be resolved.

\newpage
\section{Transfer matrix method simulations of phonon mode intensities in CrCl\texorpdfstring{$_3$}{3}\label{sec:S3}}

\begin{figure*}[h]
		\subfloat{}%
		\centering
		\includegraphics[width=0.5 \linewidth]{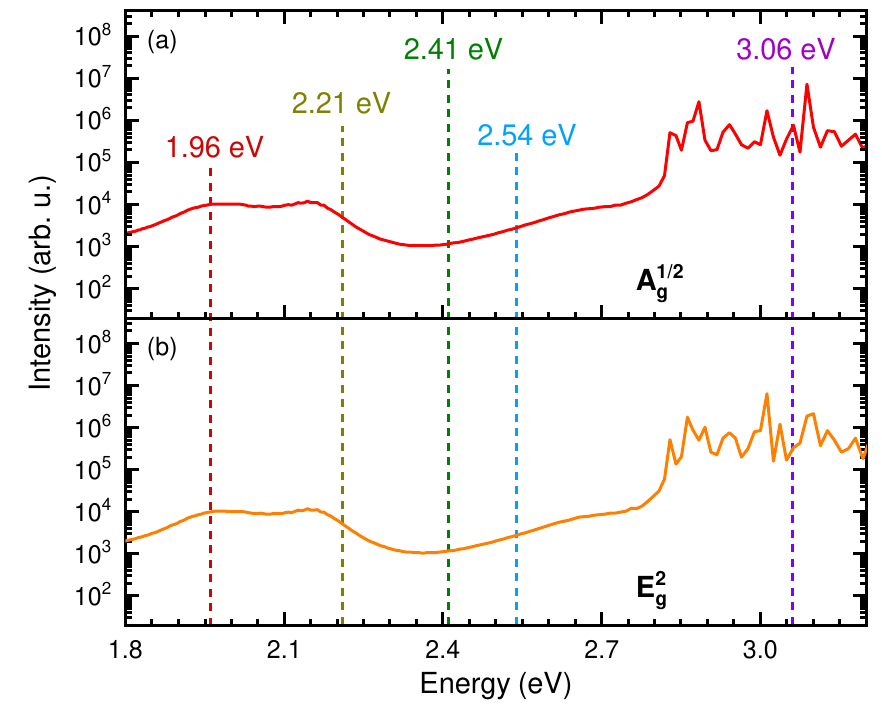}
   	\caption{Simulated enhancements of the A$^{1/2}_\textrm{g}$ and E$^{2}_\textrm{g}$ intensities using the transfer-matrix method. 
    The colored vertical dashed lines indicate the excitation energies used in the experiment.
    }
		\label{fig:modes-simulation}
\end{figure*}

Figures~\ref{fig:modes-simulation}(a) and (b) show the simulated Raman-mode intensities as a function of excitation energy for the A$^{1/2}_\textrm{g}$ and E$^{2}_\textrm{g}$ modes, respectively. 
The calculations were performed using the TMM method, described in the Methods section, at room temperature ($T$=300~K).
The simulations follow the same procedure as that applied to the A$^{3/4}_\textrm{g}$ mode (see Fig.~2(c) in the main text).

As shown in Fig.~\ref{fig:modes-simulation}, the intensity dependence on excitation energy is very similar for both modes. 
Nevertheless, subtle differences are discernible, particularly in the high-energy range (2.8--3.2~eV), where the positions of the interference extrema do not align perfectly.

Despite these minor deviations, the overall enhancement of the A$^{1/2}_\textrm{g}$ and E$^{2}_\textrm{g}$ intensities is similar to that of the A$^{3/4}_\mathrm{g}$ mode presented in Fig.~2 of the main text. 
As shown in Fig.~\ref{fig:modes-simulation}(c), the simulated A$^{1/2}_\textrm{g}$ and E$^{2}_\textrm{g}$ intensities exhibit strong enhancement for excitation energies above approximately 2.85~eV, while reaching a minimum near 2.35~eV. 
These simulations therefore confirm that 3.06~eV is the optimal excitation energy for maximizing the Raman scattering response in the investigated CrCl$_\textrm{3}$ crystal, providing strong support for the discussion of interference conditions observed for the A$^{3/4}_\mathrm{g}$ mode in the main text and enabling a more comprehensive description of the excitation-energy dependence of the Raman-mode intensities.

\newpage
\section{Angle-resolved and excitation-dependent Raman spectra of an exfoliated CrCl\texorpdfstring{$_3$}{3} flake \label{sec:S4}}

\begin{figure*}[h]
		\subfloat{}%
		\centering
		\includegraphics[width=1\linewidth]{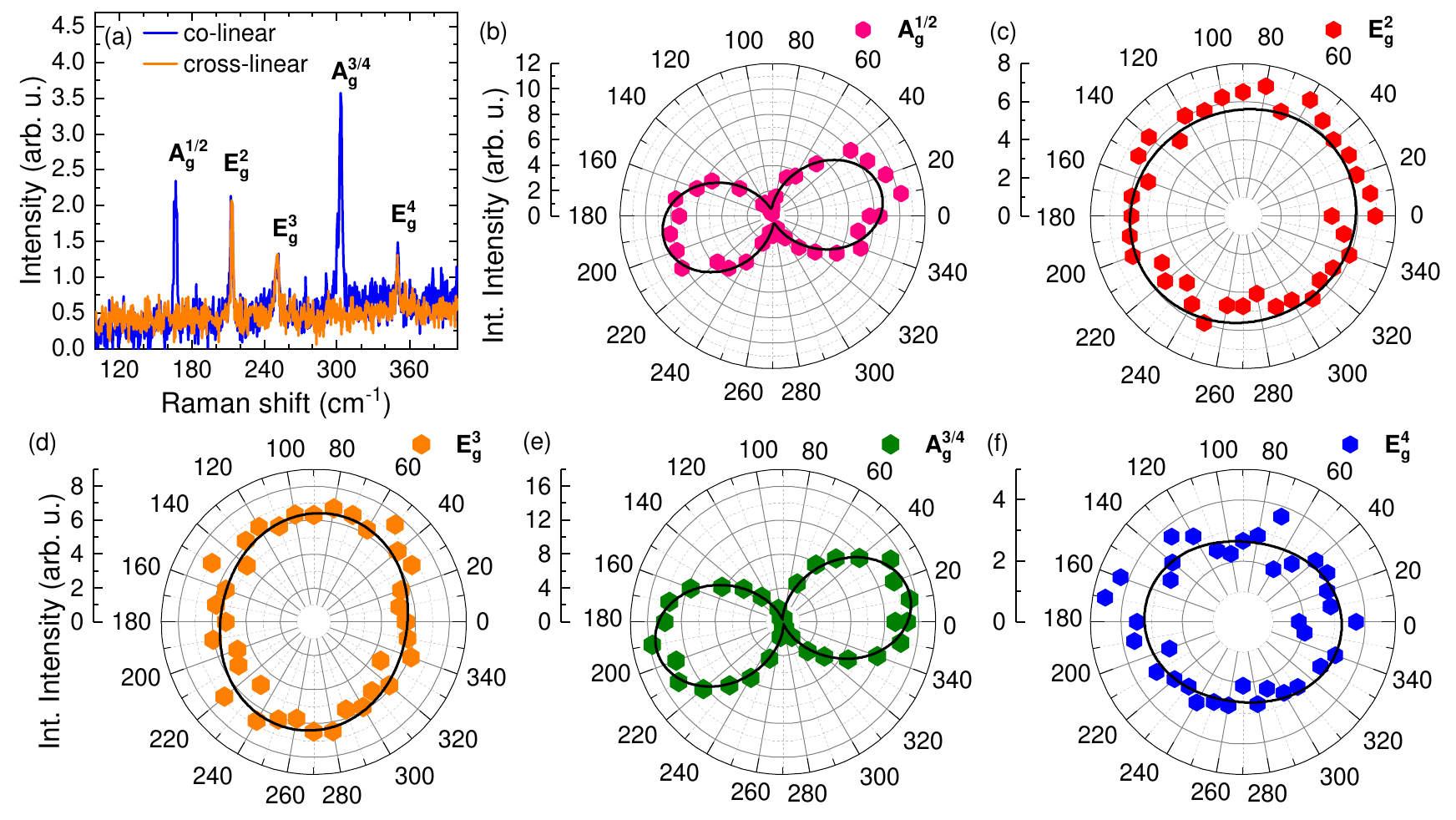}
    	\caption{(a) Low-temperature ($T$=5~K) Raman spectra of CrCl$_\textrm{3}$ flake measured in co-linear and cross-linear polarization configurations, using 3.06 eV laser excitation, with the power of 75 $\mu$W. The modes in panels (b) and (e) are a A$_\textrm{g}$ type, while the modes in panels (c), (e) and (f) are the E$_\textrm{g}$ type. The black curve represents the fitted formula using Eq.~\ref{eq:pol}.}
		\label{fig:pol-flake}
\end{figure*}

Figure~\ref{fig:pol-flake}(a) shows polarization-resolved Raman spectra of a thick exfoliated CrCl$_{\textrm{3}}$ flake with a thickness of approximately 227~nm, acquired in co-linear and cross-linear configurations (sample details are provided in the Methods section).
The spectra confirm that the thick exfoliated CrCl$_\textrm{3}$ flake exhibits the same phonon mode symmetries as the bulk crystal (see Section~\ref{sec:S2}). 
In particular, the low-frequency E$^{1}_\textrm{g}$ mode is not resolved in these measurements.

To determine the symmetry of the Raman-active modes in the exfoliated CrCl$_\textrm{3}$ flake, we measured their linear polarization dependence at low temperature ($T$=5~K).
The intensities of the Raman modes shown in Figs.~\ref{fig:polarization}(c), (d), and (e) are independent of the polarization angle. 
These modes are therefore assigned to the E$_\textrm{g}$ symmetry and correspond to in-plane atomic vibrations within the crystal lattice.
In contrast, the modes displayed in Figs.~\ref{fig:polarization}(b) and (e) exhibit a pronounced polarization dependence and their angular evolution was fitted using Eq.~(\ref{eq:pol}).
These modes are accordingly assigned to the A$_\textrm{g}$ symmetry and are associated with out-of-plane atomic vibrations.

\begin{figure*}[h]
		\subfloat{}%
		\centering
		\includegraphics[width=1 \linewidth]{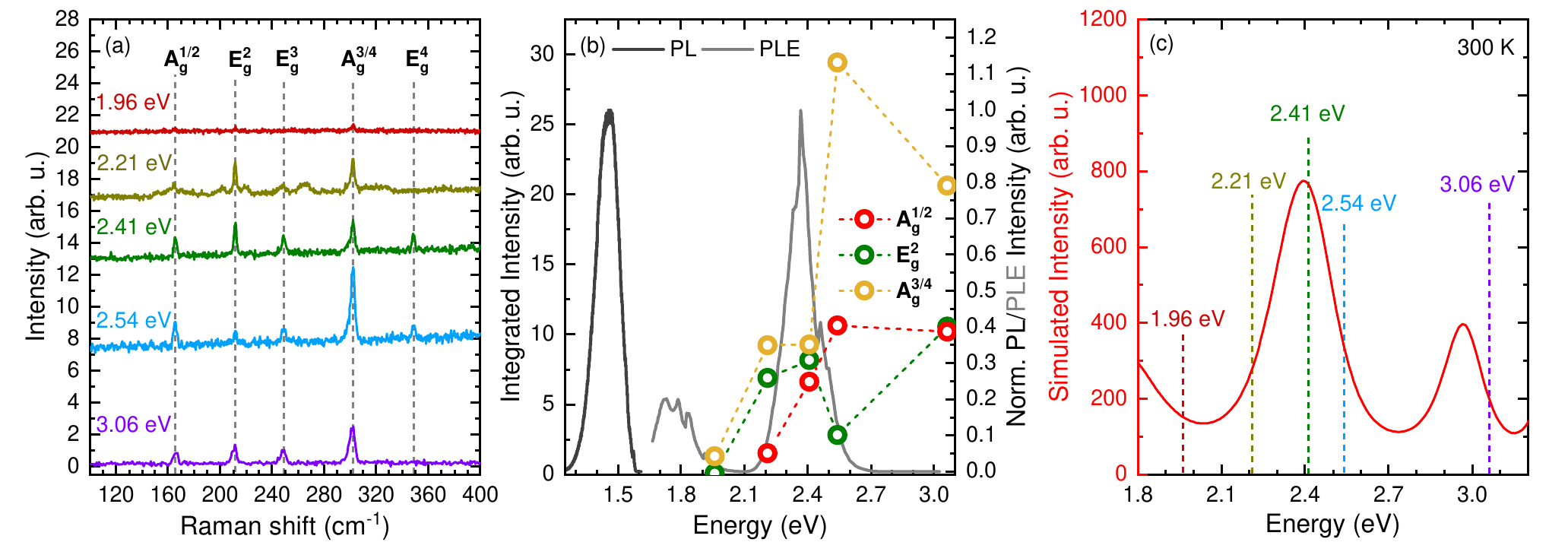}
    	\caption{(a) Raman scattering spectra of CrCl$_\textrm{3}$ thick flake exfoliated on the 285 nm SiO$_\textrm{2}$/Si substrate, measured at 5~K with different excitation energies: 1.96 eV, 2.21 eV, 2.41 eV, 2.54 eV, 3.06 eV, with an excitation power 50 $\mu$W. 
        The spectra have been vertically shifted for better visual clarity. 
        (b) Left axis: graph shows points representing the intensity of the individual Raman modes A$^{1/2}_\textrm{g}$, E$^{2}_\textrm{g}$ and A$^{3/4}_\textrm{g}$ collected with different lasers (in linear scale).
        Right axis: normalized PL (black), PLE (grey) spectra measured on the CrCl$_{3}$ flake. 
        (c) Simulated enhancement of the A$^{3/4}_\textrm{g}$ intensity using the transfer-matrix method. The colored vertical dashed lines indicate the excitation energies used in the experiment.}
		\label{fig:raman-pl-ple}
\end{figure*}

Figure~\ref{fig:raman-pl-ple}(a) presents Raman spectra of an exfoliated CrCl$_\textrm{3}$ flake acquired using excitation energies of 1.96, 2.21, 2.41, 2.54, and 3.06~eV. Although the experimental conditions were identical to those used for the CrCl$_\textrm{3}$ crystal, pronounced differences in the relative intensities of the Raman modes are observed. In contrast to the crystal results shown in Fig.~2 of the main text, the intensity variation between the 3.06~eV excitation and the other excitation energies is much less pronounced. A strong reduction of the overall Raman intensity is observed only under 1.96~eV excitation. Moreover, the relative intensities among the Raman peaks change substantially, particularly for the A$^{1/2}_{\textrm{g}}$, E$^{2}_{\textrm{g}}$, and E$^{3}_{\textrm{g}}$ modes.

To analyze the excitation-dependent intensity of the Raman modes, we deconvoluted the Raman spectra using Lorentzian functions.
The integrated intensity of the A$^{1/2}_\textrm{g}$, E$^{2}_\textrm{g}$, and A$^{3/4}_\textrm{g}$ peaks as a function of excitation energy is shown in Fig.~\ref{fig:raman-pl-ple}(b).
In contrast to the results obtained for the CrCl$_\textrm{3}$ crystal, presented in Fig.2(b) of the main article, the highest intensity of phonon modes: A$^{1/2}_\textrm{g}$ and A$^{3/4}_\textrm{g}$ is apparent under the 2.54~eV and 3.06~eV excitations, While for the E$^{2}_\textrm{g}$ mode under 2.41 eV and also 3.06 eV.
The intensities variation of the modes between the 2.54~eV and 3.06~eV excitations and the 2.41 eV and 2.21~eV ones is only about 2 times, while the intensity under 1.96~eV is substantially vanished.

To examine the resonant conditions of Raman scattering, we measured the PLE spectrum of the CrCl$_\textrm{3}$ emission, as shown in Fig.~\ref{fig:raman-pl-ple}(b). 
The PLE spectrum consists of two prominent features centered at approximately 1.8~eV and 2.4~eV. 
However, these resonances do not correlate with the excitation-energy dependence of the integrated phonon intensities. 
Consequently, the observed Raman enhancement cannot be understood in terms of conventional electron–phonon coupling leading to resonant Raman scattering.

As the PLE measurements, similarly to those for the bulk crystal discussed in the main text, do not sufficiently account for the enhancement profiles of the phonon modes in the exfoliated CrCl$_\textrm{3}$ flake, we simulated the excitation-energy dependence of the A$^{3/4}_\textrm{g}$ intensity. 
The simulations were performed using the transfer-matrix method (TMM), following the same procedure as for the CrCl$_\textrm{3}$ crystal in the main text, and are shown in Fig.~\ref{fig:raman-pl-ple}(c). 
The simulated spectrum consists of two enhancement features centered at approximately 2.40~eV and 2.95~eV, with an overall enhancement factor of about four between the maxima and minima. 
According to the simulation, the 2.41~eV excitation should yield the strongest Raman signal, whereas experimentally the highest intensities are observed under 2.54~eV and 3.06~eV excitation. 
Nevertheless, the lowest Raman intensity is predicted for 1.96~eV excitation, in agreement with the experimental results.

Because the simulations were carried out at 300~K, while the Raman measurements were performed at 5~K, the discrepancy between experiment and theory can be attributed to temperature effects.
In particular, as discussed in Sec.~\ref{sec:S6}, the integrated phonon intensities increase by approximately a factor of two when the temperature is raised from 5 to 300~K, which is comparable to the difference between the minimum and maximum values near the simulated enhancement peak at $\sim$2.95~eV. 
Therefore, the deviation between the experimentally observed excitation-energy dependence of the phonon intensities and the simulated enhancement spectrum can be explained by the temperature-induced modifications of the electronic structure and phonon energies in exfoliated CrCl$_\textrm{3}$.

\newpage
\section{Temperature evolutions of the Raman mode linewidths and intensities in CrCl\texorpdfstring{$_3$}{3} crystals.\label{sec:S5}}

\begin{figure}[h]
		\subfloat{}%
		\centering
		\includegraphics[width=0.7 \linewidth]{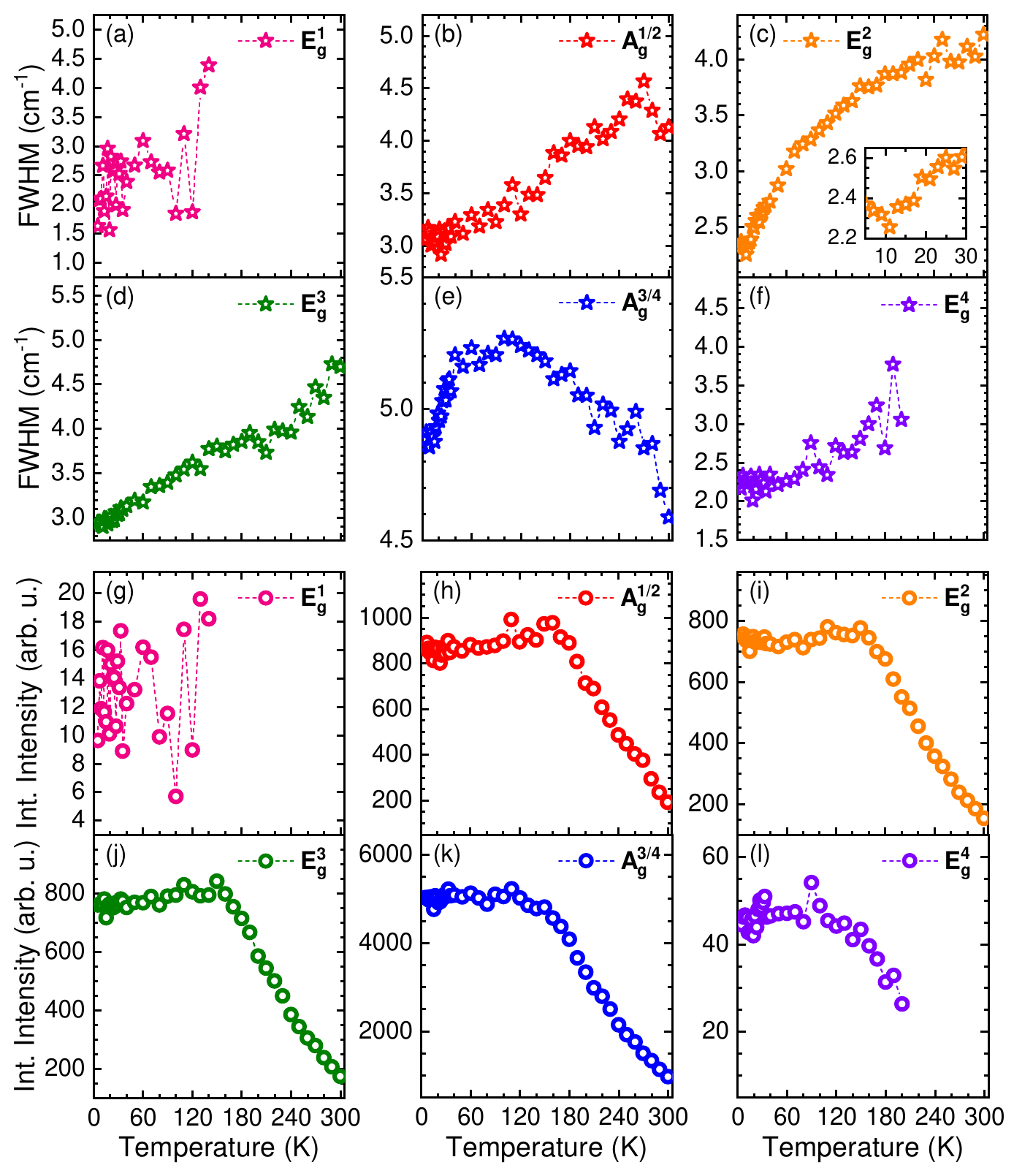}
    	\caption{Temperature dependencies of (a)–-(f) the linewidths (full width at half maximum, FWHM) and (g)–-(l) the integrated intensities of all phonon modes observed in the Raman spectra shown in Fig.~3 of the main text.}
		\label{fig:intensity-fwhm}
\end{figure}

Figure~\ref{fig:intensity-fwhm} presents the temperature dependence of the Raman modes observed in a CrCl$_\textrm{3}$ crystal over the temperature range of 5–300~K. 
Panels (a)-–(f) show the temperature evolution of the full width at half-maximum (FWHM) of the phonon modes, while panels (g)–-(l) display the corresponding evolution of their integrated intensities. 
In the following, we discuss in detail the temperature-dependent behaviors of both the FWHMs and the integrated intensities for all observed phonon modes.

With increasing temperature, the FWHMs of all E$_\textrm{g}$ modes, $i.e.$, E$^{1}_{\textrm{g}}$, E$^{2}_{\textrm{g}}$, E$^{3}_{\textrm{g}}$, and E$^{4}_{\textrm{g}}$, exhibit a clear monotonic increase, see Figs.~\ref{fig:intensity-fwhm}(a), (c), (d), and (f).
Such behavior is typical and can be attributed to lattice anharmonicity, arising from the combined effects of thermal expansion of the crystal lattice and enhanced phonon–phonon scattering processes.\cite{Balkanski1983, Menendez1984, Taube_2015, Joshi_2016, Sarkar_2020} 
Subtle differences in the temperature evolution among the individual modes are also observed.

The behavior of the E$^{2}_{\textrm{g}}$ mode, shown in Fig.~\ref{fig:intensity-fwhm}(c), is distinct from that of the other modes. 
Although the overall trend indicates thermal broadening, the low-temperature regime (5–30~K), highlighted in the inset, reveals an anomalous behavior. 
Specifically, the FWHM initially decreases upon warming, reaching a minimum at approximately 10~K, before increasing with further temperature growth. 
Notably, this minimum coincides with the proposed magnetic phase transition from the antiferromagnetic to a domain-like ferromagnetic order at the N\'eel temperature ($T_\mathrm{N}$$\approx$14~K).

The temperature dependencies of both A$_\textrm{g}$ modes, $i.e.$, A$^{1/2}_{\textrm{g}}$ and A$^{3/4}_{\textrm{g}}$, shown in Figs.~\ref{fig:intensity-fwhm}(b) and (e), are the most challenging to interpret.
This difficulty arises because each Raman feature consists of two closely spaced modes that cannot be resolved under 3.06~eV excitation. 
Nevertheless, while the A$^{1/2}_{\textrm{g}}$ mode exhibits a temperature evolution similar to that of the E$_\textrm{g}$ phonons, the behavior of A$^{3/4}_{\textrm{g}}$ is particularly intriguing. 
Its linewidth initially increases with temperature, reaching a maximum near 80~K, and subsequently exhibits an anomalous decrease. 
This behavior cannot be unambiguously assigned, as it may originate either from a structural phase transition from the $\bar{R}3$ to the $C2/m$ phase,\cite{McGuire_2017, Lis_2025} or from the dual-peak nature of the mode itself: with increasing temperature, one component progressively weakens, leaving a single observable mode at room temperature.

The thermal evolutions of the integrated intensities of the observed phonon modes exhibit much smaller variations among the modes, as shown in Figs.~\ref{fig:intensity-fwhm}(g)–(l). 
With the exception of the E$^{1}_{\textrm{g}}$ mode [Fig.~\ref{fig:intensity-fwhm}(g)], which displays noticeable scatter likely due to its weak Raman intensity, the remaining modes, $i.e.$, E$^{2}_{\textrm{g}}$, E$^{3}_{\textrm{g}}$, E$^{4}_{\textrm{g}}$, A$^{1/2}_{\textrm{g}}$, and A$^{3/4}_{\textrm{g}}$, exhibit very similar temperature dependencies. 
Their integrated intensities are only weakly affected by increasing temperature up to approximately 160~K, with modest enhancements observed for the A$^{1/2}_{\textrm{g}}$, E$^{2}_{\textrm{g}}$, and E$^{3}_{\textrm{g}}$ modes. 
At higher temperatures, the integrated intensities decrease markedly, resulting in values approximately 5–8 times smaller at room temperature compared to those in the 5–160~K range. 
We attribute this pronounced reduction in intensity to the structural phase transition from the $\bar{R}3$ to the $C2/m$ phase,\cite{McGuire_2017, Lis_2025} which may also induce concomitant changes in the electronic band structure of the CrCl$_\textrm{3}$ crystal.

\newpage
\section{Temperature-dependent Raman spectra of an exfoliated CrCl\texorpdfstring{$_3$}{3} flake. \label{sec:S6}}

\begin{figure*}[h]
		\subfloat{}%
		\centering
		\includegraphics[width=1 \linewidth]{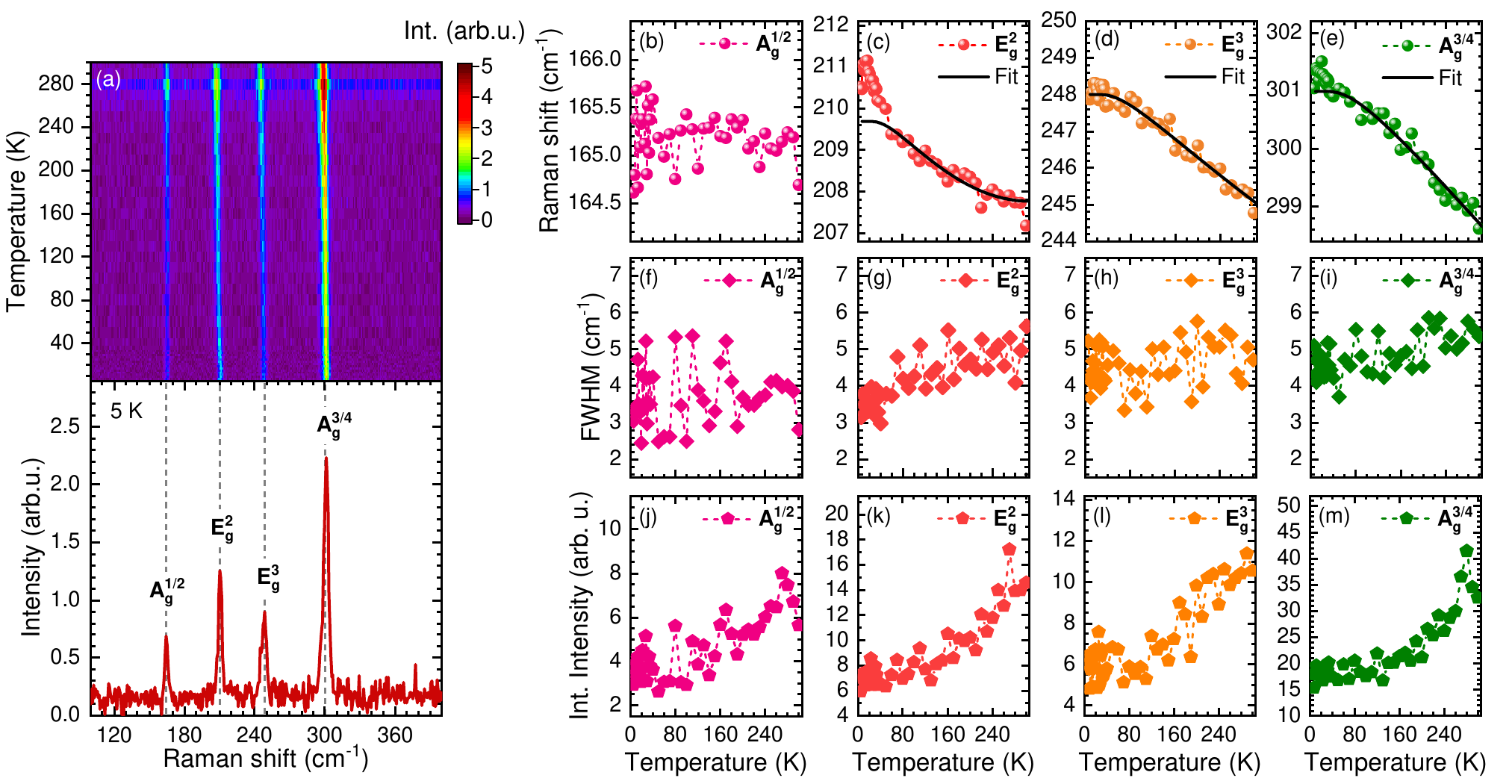}
    	\caption{
        (a) The top panel shows a false-color map of the Raman spectra of a CrCl$_\textrm{3}$ crystal, while the bottom panel presents the Raman spectrum measured at $T$~=~5~K.
        Temperature dependence of (b)–(e) the Raman shifts, (f)–(i) the FWHMs, and (j)–(m) the integrated intensities of all observed phonon modes. 
        The solid black curves in panels (b)–(e) represent fits using Eq.~1 described in the main text. }
		\label{fig:temp-flake}
\end{figure*}

Figure~\ref{fig:temp-flake} presents temperature-dependent Raman measurements performed on an exfoliated CrCl$_\textrm{3}$ flake.
Panel (a) shows a false-color intensity map illustrating the evolution of the Raman spectrum over the temperature range from 5~K to 300~K, together with a representative spectrum acquired at 5~K.
Within the investigated spectral window, four prominent phonon modes are clearly resolved: A$^{1/2}_\textrm{g}$ ($\sim$166~cm$^{-1}$), E$^{2}_\textrm{g}$ ($\sim$210~cm$^{-1}$), E$^{3}_\textrm{g}$ ($\sim$248~cm$^{-1}$), and A$^{3/4}_\textrm{g}$ ($\sim$301~cm$^{-1}$).

Panels (b)–(e) of Fig.~\ref{fig:temp-flake} display the temperature dependences of the Raman shifts of the four observed phonon modes, extracted using Lorentzian fits to the Raman spectra.
In general, the observed trends are similar to those shown in Fig.~3 of the main text for the CrCl$_\textrm{3}$ crystal.
The experimental data were fitted using the Balkanski formula, given in Eq.~1 of the main text.
All modes exhibit a monotonic redshift with increasing temperature, which is primarily attributed to the lattice anharmonicity arising from the combined effects of thermal expansion of the crystal lattice and enhanced phonon–phonon scattering processes.\cite{Balkanski1983, Menendez1984}

The A$^{1/2}_\textrm{g}$ mode is significantly weaker than that measured in the bulk crystal, resulting in noticeable fluctuations of its Raman shift as a function of temperature (see Fig.~\ref{fig:temp-flake}(b)).
Nevertheless, the characteristic redshift of the A$^{1/2}_\textrm{g}$ mode with increasing temperature is still observed.
The temperature dependences of the E$^{2}_\textrm{g}$ and A$^{3/4}_\textrm{g}$ modes, shown in Figs.~\ref{fig:temp-flake}(c) and (e), respectively, are particularly notable and closely mirror the behavior observed in the bulk crystal.
In the low-temperature regime (5–30~K), both the E$^{2}_\textrm{g}$ and A$^{3/4}_\textrm{g}$ modes exhibit an increase in Raman shift, reaching a maximum at 10~K, followed by a redshift at higher temperatures.
As discussed in the main text, the anomaly at 10~K can be attributed to the proximity to the N{\'e}el temperature ($T_\mathrm{N}$ = 14~K) and is associated with the transition from the antiferromagnetic phase to a domain-like ferromagnetic phase.
In contrast, the temperature evolution of the E$^{3}_\textrm{g}$ mode shows no detectable influence of spin–phonon coupling, as its behavior over the entire temperature range can be well described by the Balkanski model.

The temperature dependencies of the FWHMs for all four Raman peaks, shown in Figs.~\ref{fig:temp-flake}(f)–(i), exhibit substantial scatter. 
This effect is particularly pronounced for the A$^{1/2}_{\textrm{g}}$ mode, whose FWHM varies by more than a factor of two between consecutive temperature points separated by 20~K. 
For the remaining phonon modes, $i.e.$, E$^{2}_{\textrm{g}}$, E$^{3}_{\textrm{g}}$, and A$^{3/4}_{\textrm{g}}$, an overall monotonic increase in FWHM with temperature can still be discerned, consistent with standard anharmonic lattice effects and enhanced electron–phonon interactions.
However, this behavior lacks the specific features observed in the analogous analysis of the CrCl$_\textrm{3}$ crystal, as discussed in Sec.~\ref{sec:S5}.

The influence of temperature on the integrated intensities of the phonon modes, shown in Figs.~\ref{fig:temp-flake}(j)–(m), is similar for all modes, $i.e.$, their intensities increase by approximately a factor of two upon warming from 5 to 300~K. 
This behavior is markedly different from that observed for the CrCl$_\textrm{3}$ crystal, presented in Figs.~\ref{fig:intensity-fwhm}(g)–(l). 
As shown in Fig.~\ref{fig:raman-pl-ple}(c), the 3.06~eV excitation lies on the slope of the enhancement peak centered at approximately 2.95~eV. Simulations indicate that the minimum and maximum intensities in the vicinity of this peak can differ by about a factor of two, which may account for the observed variation in phonon intensities. 
Consequently, given that temperature modulates both the electronic structure and the phonon energies, and considering possible quantitative discrepancies between experimental results and theoretical calculations, we primarily attribute the observed temperature evolution to the interference effect discussed in Sec.~\ref{sec:S4}.

\clearpage

\end{document}